\documentclass[aps,prd,showkeys,nofootinbib]{revtex4-2}

\usepackage{amsmath}
\usepackage{amssymb}
\usepackage{hyperref}
\usepackage{xcolor}
\usepackage{graphicx}
\usepackage{dcolumn}
\usepackage[loop]{animate}
\usepackage{comment}
\usepackage[normalem]{ulem}
\usepackage[bottom]{footmisc}
\usepackage{float}

\newcommand{\bs}{\boldsymbol}
\newcommand{\p}{\partial}

\newcommand{\ptnabla}{{}^\perp\tilde{\nabla}}
\newcommand{\divv}{\text{div}}

\def\la{\; \raise0.3ex\hbox{$<$\kern-0.75em\raise-1.1ex\hbox{$\sim$}}\;}
\def\ga{\;  \raise0.3ex\hbox{$>$\kern-0.75em\raise-1.1ex\hbox{$\sim$}}\;}

\begin{document}

\title{
Role of particle diffusion in shaping the gravitational wave signal \\
from neutron star inspirals
}

\date{\today}

\author{Elena M. Kantor}
\author{Mikhail E. Gusakov}
\author{Kirill Y. Kraav}
\affiliation{Ioffe Institute, Polytekhnicheskaya 26, St.-Petersburg 194021, Russia}


\begin{abstract}

It is commonly believed that the dissipative properties of superdense matter play a negligible role in modeling gravitational waveforms from neutron star inspirals. This study aims to investigate whether this presumption holds true for the often neglected dissipative process associated with particle diffusion in superconducting neutron stars. As we demonstrate, diffusion effects can significantly impact the phase of the gravitational wave from the inspiral, manifesting at a magnitude of a few tens of milliradians at large orbit separations, equivalent to orbital frequencies of a few hertz. We also find that dissipation resulting from particle diffusion might increase the neutron star's temperature to approximately $10^7\rm K$ during the inspiral.

\end{abstract}


\maketitle

 
\section{Introduction}

Neutron stars (NS) are unique astrophysical objects containing matter at densities surpassing those found within atomic nuclei. This characteristic makes NSs ideal laboratories for investigating the properties of superdense matter, shedding light on various aspects of fundamental physics. The recent detection of the gravitational wave (GW) signal from the binary NS inspiral in 2017 (GW170817) by LIGO and VIRGO detectors \cite{GW17} has uncovered exciting new possibilities for studying and understanding NSs, significantly amplifying the role of neutron stars in modern observational astrophysics.

In the first approximation, GW emitted by inspiralling NSs can be modelled by ignoring their finite sizes and using the approximation of point masses. The GW signal in this approximation does not carry any information about NS interiors and depends only on the chirp mass of the binary. To study the properties of the stellar matter, one has to relax the point-mass approximation and account for the finite NS sizes and their distortion by tidal forces. The distortion of the star increases gravitational radiation from the binary, which starts to depend on the tidal deformability of the NS (on the NS response to the quadrupole tidal perturbation) \cite{fh08}. As a result, since the GW energy is radiated at the expense of orbital energy, distortion accelerates the inspiral, introducing observable corrections to the point mass model of the signal. The observed GW signal from GW170817 has already allowed one to impose constraints on these corrections and, consequently, on the tidal deformability value, see, e.g., Ref.\ \cite{constraint19}. Numerous more subtle effects (e.g., \cite{dhs21}) are most probably beyond the sensitivity of currently operating detectors but can be within reach of the next generation of GW detectors (such as the Einstein Telescope, Cosmic Explorer, or LISA), which are expected to have higher sensitivity 
and an operating frequency range extending, in particular, to lower frequencies.

Thus, the correct interpretation of future observations requires accurate models of inspiralling NSs, accounting for the influence of various effects on the GW signal.
A lot of work has been done in this direction, see, e.g., the review \cite{dhs21} and references therein. In particular, various dissipative mechanisms in NS matter may dissipate the distortion energy, eventually transforming orbital energy into heat and accelerating the inspiral.
While the influence of viscous dissipation has already been addressed to some extent (see, e.g., Refs.\ \cite{Lai94,yw17,aw19,kksk22,most22}), the effect of particle diffusion \cite{braginskii65,ys91,ys91b,dgs20,dg21,kgk21,gg23}, to the best of our knowledge, has never been considered in this context. In the present work, we quantify its importance for GW signal during NS inspirals. 

The paper is organized as follows. Section \ref{theory} introduces the formalism and discusses the equations to be solved. Section \ref{results} contains our numerical results, Section \ref{disc} presents discussion and summary. The paper also has a number of extensive appendices.
In Appendices \ref{AppA} and \ref{AppB}, the formalism of Section \ref{theory} is developed and justified in detail. Appendix \ref{AppC} describes binary evolution in the Newtonian limit. Appendix \ref{AppD} presents the effect of diffusion on the stellar perturbation. 
Finally, Appendix \ref{AppE} presents plots of various quantities relevant to the discussion in the main text.



\section{Theoretical framework}\label{theory}


\subsection{Perturbation equations of a neutron star in a binary system}
\label{perturb} 

In what follows, we use the unit system where the speed of light $c=1$ and Boltzmann constant $k_{\rm B}=1$. Let us consider a perturbation of a non-rotating neutron star in a binary system, induced by its companion. In this study, we assume that the matter of the perturbed star is composed of strongly superconducting protons $(p)$ (superconducting component) and of normal (i.e., nonsuperfluid) neutrons $(n)$, electrons $(e)$ and muons $(\mu)$ (normal component). We characterize such matter by the pressure $P$, energy density $\varepsilon$ and temperature $T$, and each of the particle species $k=(n,p,e,\mu)$ by the corresponding number densities $\{n_k\}$, chemical potentials $\{\mu_k\}$ and electrical charges $\{e_k\}$. From now on we use Latin indices $m$, $k$, $l$, $q$, and $r$ to label different particle species. In the absence of diffusion the particle current densities of different particle species equal \cite{dg21}
\begin{gather}
j_{(n),(e),(\mu)}^\mu=n_{n, e, \mu}u^\mu , \qquad j_{(p)}^\mu=n_{p} u^\mu+Y_{pp} w_{(p)}^\mu, 
\end{gather}
where $u^\mu$ is the collective four-velocity of the normal component, $Y_{pp}$ is the relativistic analogue of the superfluid proton number density, and the vector $w_{(p)}^\mu$ describes the relative motion of normal and superconducting components 
\footnote{In the Newtonian limit $Y_{pp}{\bs w}_{(p)}=n_s({\bs u}_s-{\bs u})$, 
where $n_s$ and ${\bs u}_s$
are the superfluid proton number density and superfluid velocity, respectively.}
(from now on we use Greek indices for tensor components).
Diffusion introduces additional corrections to these currents. In this study, we restrict our consideration to the limit of slow diffusion by assuming that these corrections are small
(the validity of this approximation for the considered problem will be discussed in detail  
in Sec.\, \ref{validity} and Appendix \ref{AppE}).
As shown in Appendix \ref{AppA}, in these circumstances one can present the expressions for $j_{(k)}^\mu$ as (we assume summation over repeated Greek and Latin indices)
\begin{gather}
j_{(k)}^\mu=n_k \tilde{u}^\mu+H_{km}\Delta j_{(m)}^\mu, \qquad \tilde{u}^\mu=u^\mu+\frac{\mu_p Y_{pp}}{P+\varepsilon}w_{(p)}^\mu, \qquad \tilde{u}_\mu \tilde{u}^\mu=-1, \\
\Delta j_{(m)}^\mu=-\frac{D_{mk}H_{lk}}{T}\ptnabla^\mu\mu_l, \quad \ptnabla^\mu=(g^{\mu\nu}+\tilde{u}^\mu\tilde u^\nu)\nabla_\nu, \quad H_{mk}=\delta_{mk}-\frac{e_k}{e_{p}}\biggl(\delta_{mp}-\frac{\mu_p n_m}{P+\varepsilon}\biggr), 
\end{gather}
where $g_{\mu\nu}$ is the metric tensor, $\nabla_\nu$ is the covariant derivative, and $D_{mk}$ is the non-negative-definite matrix of diffusion coefficients, introduced in Ref.\ \cite{dgs20}. Generally, $D_{mk}$ can be presented as a function of temperature $T$, number densities $n_l$ and chemical potentials $\mu_l$ of all particle species. Further, in the limit of slow diffusion the standard thermodynamic relations of nonsuperconducting degenerate matter hold true, while the stress-energy tensor $T^{\mu\nu}$ resembles that of a perfect nonsuperconducting fluid:
\begin{gather}
d\varepsilon=\mu_k dn_k, \qquad dP=n_kd\mu_k, \qquad T^{\mu\nu}=(P+\epsilon)\tilde{u}^\mu \tilde{u}^\nu + P g^{\mu\nu}.
\end{gather}
As discussed in Appendix \ref{AppA}, the four-velocity $\tilde{u}^\mu$ multiplied by $(P+\varepsilon)$ has a meaning of the momentum density of the fluid. To describe the gravitational field in a neutron star in a binary system, we employ the inertial reference frame, whose origin each moment of time coincides with the neutron star center of mass. In this frame the metric tensor can be approximately written, in the spherical coordinates $x^\mu=(t,r,\theta,\varphi)$, as (see Appendix \ref{AppB})
\begin{gather}
\label{gtensor}
ds^2=g_{\mu\nu}dx^\mu dx^\nu=-e^{\nu_0+2U}dt^2+e^{\lambda_0}dr^2+r^2(d\theta^2+\sin^2\theta d\varphi^2),
\end{gather}
where $\nu_0(r)$ and $\lambda_0(r)$ are the standard metric functions, which describe the equilibrium spacetime of an isolated spherically symmetric neutron star, and $U(t,r,\theta,\varphi)$ is the gravitational potential, induced in the star by its companion. It can be represented as a sum $U=U_0+\tilde{U}$. Here the term $U_0$ produces the uniform acceleration throughout the star and, therefore, cannot lead to stellar deformations. The term $\tilde{U}$, further referred to as the {\it tidal gravitational potential}, produces coordinate-dependent acceleration and, therefore, is responsible for the stellar deformations. Explicit expressions for $U$ and $\tilde U$ are not important here and will be specified later in Sec.\ \ref{sec24}. The potential $U$ (and $\tilde{U}$) is weak compared to the local gravitational field and will be treated further as a small perturbation.

Now we are in position to write down the linearized perturbed equations for a neutron star in a close binary. Let us denote the Eulerian perturbation of a quantity $f$ as $\delta f=f-f_0$, where $f_0$ is its equilibrium value (in the system with $U=0$). For simplicity, in what follows, we adopt the Cowling approximation, i.e., we ignore the perturbations of the gravitational field, $\delta g_{\mu\nu}=0$, so that the metric tensor in the perturbed star is still given by Eq.\ \eqref{gtensor}. We also introduce the Lagrangian displacement ${\bs \xi}$ by $\delta\tilde{\bs u}=e^{-\nu_0/2}\dot{{\bs \xi}}$, where the dot denotes the derivative $\p/\p t$ [note that due to linearized normalization condition, $\delta(\tilde{u}_\mu \tilde{u}^\mu)=0$, we have $\delta\tilde{u}^t=\xi^t=0$]. Then, small perturbations of a neutron star are governed by the following equations (see also Appendices \ref{AppA} and \ref{AppB}):
\begin{gather}
\left\{
\begin{gathered}
e^{-\nu_0}(P_0+\varepsilon_0)\ddot{\bs\xi}=-{\pmb\nabla}\delta P+\frac{\delta P+\delta\varepsilon}{P_0+\varepsilon_0}{\pmb\nabla} P_0-(P_0+\varepsilon_0){\pmb\nabla} \tilde{U} \\
\delta \dot{n}_k+\divv(n_{k0}{\dot{\bs\xi}})=-\divv(H_{km0}\Delta {\bs j}_m e^{\nu_0/2}),
\\
\delta \varepsilon=\mu_{k0}\delta n_k, \quad \delta P=(\p P/\p n_k)_0\delta n_k=n_{k0}\delta\mu_k, \quad \delta\mu_k=(\p\mu_k/\p n_m)_0\delta n_m
\end{gathered}
\right..
\label{eu}
\end{gather}
The first equation in (\ref{eu}) is the linearized Euler equation, $\delta[(g^{\rho\nu}+\tilde{u}^\rho \tilde{u}^\nu)\nabla_\mu T^{\mu}_\nu]=0$; the second is the set of linearized continuity equations, $\delta[\nabla_\mu j_{(k)}^\mu]=0$, with $\Delta {\bs j}_m$ given by Eq.\ \ref{jmu5}. 
The particle source in continuity equations, caused by the deviation from the beta-equilibrium, is negligible due to strong proton superconductivity. The remaining equations follow from the thermodynamic relations.  Physical solutions of these equations should be regular at the stellar center and correspond to vanishing total pressure at the stellar surface. The latter condition is equivalent to requiring that the Lagrangian pressure perturbation $\Delta P=\delta P+({\bs\xi}\cdot{\pmb\nabla})P_0=0$ at the surface. In what follows, we will assume that diffusion coefficients $D_{km}$ and their derivatives vanish at the surface. Then from the continuity equations and thermodynamic relations it follows that $\Delta P|_{r=R}=-\gamma P_0 \divv{\bs\xi}|_{r=R}=0$, where $\gamma P_0=n_{m0}(\p P/\p n_m)_0$, and $R$ is the stellar radius. Explicit definitions of the three-dimensional operators $\pmb\nabla$, ``$\divv$'', and of scalar product $(\cdot)$ in curvilinear coordinates are given in Appendix \ref{AppB}.


\subsection{Energy balance equation}

Let us decompose each thermodynamic perturbation as $\delta f=\delta f_{\rm ad}+\delta f_{\rm diff}$, where $\delta f_{\rm ad}$ is the ``adiabatic'' contribution, which is formally given by the same expression as in the absence of diffusion, and $\delta f_{\rm diff}$ is the diffusive correction. From the linearized continuity equations we see that
\begin{gather}
\delta n_{k, {\rm ad}}=-\divv(n_{k0}{\bs\xi}), \qquad \delta n_{k, {\rm diff}}=-\int\limits_{-\infty}^t \divv(H_{km0}\Delta {\bs j}_m e^{\nu_0/2}) dt.
\end{gather}
Therefore, for any thermodynamic quantity $f$ the adiabatic and diffusive contributions to $\delta f$ can be found as 
\begin{gather}
\delta f_{\rm ad}=\biggl(\frac{\p f}{\p n_k}\biggr)_0 \delta n_{k,{\rm ad}}=-\biggl(\frac{\p f}{\p n_k}\biggr)_0\divv(n_{k0}{\bs\xi}), \\ 
\delta f_{\rm diff}=\biggl(\frac{\p f}{\p n_k}\biggr)_0 \delta n_{k,{\rm diff}}=-\biggl(\frac{\p f}{\p n_k}\biggr)_0\int\limits_{-\infty}^t \divv(H_{km0}\Delta {\bs j}_m e^{\nu_0/2}) dt.
\end{gather}
Let us insert such decompositions for $\delta P$ and $\delta \varepsilon$ in the linearized Euler equation. Notably, one can show that $\delta \varepsilon_{\rm diff}=0$ (see Appendix \ref{AppB}), and the only diffusive contribution in this equation comes from $\delta P_{\rm diff}$. Rearranging the terms and using one of the Tolman-Oppenheimer-Volkoff equations, ${\pmb\nabla} P_0=-(P_0+\varepsilon_0){\pmb\nabla}(\nu_0/2)$ \cite{hpy07}, we rewrite the Euler equation as [see \eqref{Lxi} in Appendix \ref{AppB} for the explicit form of the operator $\hat{\mathcal{L}}$]
\begin{gather}
e^{-\nu_0/2}(P_0+\varepsilon_0)\ddot{\bs\xi}-\hat{\mathcal{L}}{\bs\xi}=-{\pmb\nabla}(\delta P_{\rm diff} e^{\nu_0/2})-e^{\nu_0/2}(P_0+\varepsilon_0){\pmb\nabla}\tilde{U}.
\label{eu2}
\end{gather}
The symmetry property \eqref{symmetric} of $\hat{\mathcal{L}}$ allows one to derive the energy conservation law from (\ref{eu2}). To demonstrate that, let us first consider the simplest case, when diffusion and tidal potential are absent. Then, multiplying the Euler equation (\ref{eu2}) by $\dot{{\bs\xi}}$, integrating it over the stellar volume and using \eqref{symmetric}, we find 
\begin{gather}
\label{Emech ad}
\frac{d E_{\rm mech, \, ad}}{dt}=0, \qquad
E_{\rm mech, \, ad}(\bs \xi)\equiv\int\biggl[\frac{1}{2}e^{-\nu_0/2}(P_0+\varepsilon_0)\dot{{\bs\xi}}^2-\frac{1}{2}{\bs\xi}\hat{\mathcal{L}}{\bs\xi}\biggr]dV.
\end{gather}
This equation describes the conservation of the total mechanical energy $E_{\rm mech, \, ad}$ of a perturbation in the absence of external forces and dissipative effects. The first term on the left-hand side can be interpreted as the kinetic energy of the perturbation, while the second -- as its potential energy. Now, accounting for diffusion and tidal potential and operating in the same manner, we obtain the generalized energy conservation law [see \eqref{Dxi2} in Appendix \ref{AppB} for the explicit form of the operator $\hat{\mathcal{D}}$]:
\begin{gather}
\label{Ebalance}
\frac{d E_{\rm mech}}{dt}=\dot{E}_{\rm diff}+A_U, \\
E_{\rm mech}\equiv \int \biggl[\frac{1}{2}e^{-\nu_0/2}(P_0+\varepsilon_0)\dot{{\bs\xi}}^2-\frac{1}{2}{\bs \xi}\hat{\mathcal{L}}{\bs\xi}+({\bs\xi}\cdot{\pmb\nabla})(\delta P_{\rm diff} e^{\nu_0/2})\biggr]dV, \label{Emech} \\
\dot{E}_{\rm diff}\equiv\int {\bs\xi}\hat{\mathcal{D}}{\bs\xi} dV, \quad
A_U\equiv-\int e^{\nu_0/2}(P_0+\varepsilon_0)(\dot{{\bs\xi}}\cdot{\pmb\nabla})\tilde{U} dV.
\end{gather}
As discussed in Appendix \ref{AppB}, the operator $\hat{\mathcal{D}}$ satisfies the condition $\int dV {\bs\xi}\hat{\mathcal{D}}{\bs\xi} \leq 0$ for any $\bs\xi$. This property allows one to interpret the first term on the right-hand side of the equation \eqref{Ebalance} as the mechanical energy loss rate through diffusion, $\dot E_{\rm diff}$, since the latter should by its definition be non-positive. Next, the second term on the right-hand side of \eqref{Ebalance} describes the work $A_U$ done by the tidal forces on the star per unit time. Finally, the third term on the right-hand side of \eqref{Emech} describes the correction to the mechanical energy, introduced by diffusive currents. Thus, equation \eqref{Ebalance} has a very clear physical interpretation: the change $\dot{E}_{\rm mech}$ of the mechanical energy $E_{\rm mech}$ is caused by the energy dissipation through diffusion and the work done by tidal forces.  

For the future discussion, we would like to note that the obtained energy loss rate coincides with that derived
in Ref.\ \cite{kgk23}, and can be written as (see Appendix \ref{AppB}):
\begin{gather}
\label{EdiffMu}
\dot{E}_{\rm diff}=\int {\bs\xi}\hat{\mathcal{D}}{\bs\xi} dV  =-\int \frac{1}{T}H_{qk0}H_{lm0}D_{km0} ({\pmb\nabla}\delta\mu_q^\infty) \cdot ({\pmb\nabla}\delta\mu_l^\infty) dV=\int T (\Delta{\bs j}_k \cdot {\bs d}_{k}) e^{\nu_0} dV, \quad {\bs d}_k=H_{lk0}{\pmb\nabla}\biggl(\frac{\delta\mu_l}{T}\biggr),
\end{gather}
where the superscript $^\infty$ introduces gravitational redshift ${e}^{\nu_0/2}$ to the corresponding quantity. From this formula, which also follows from the general results of Refs.\ \cite{dgs20,dg21}, one sees that larger chemical potential perturbations lead to stronger dissipation. This formula also implies that stronger deviations of the neutron star from diffusive equilibrium lead to stronger dissipation. Indeed, the farther the star is from diffusive equilibrium, the larger the diffusive currents $\Delta{\bs j}_k$, aiming to restore the equilibrium. In particular, if the star is close to diffusive equilibrium, currents are weak and $\dot{E}_{\rm diff}$ is suppressed. For the future discussion, we would also like to note that, according to Ref.\ \cite{kgk21}, $\dot{E}_{\rm diff}$ can be equivalently written in terms of chemical potential imbalances, describing the deviation of the matter from $\beta$-equilibrium. As a result, if chemical potential imbalances are weakly perturbed, $\dot{E}_{\rm diff}$ is suppressed.


\subsection{The effect of diffusion on the evolution of a binary system}\label{IIC}

One effect diffusion brings to the problem is that the neutron star is heated at the rate $(-\dot{E}_{\rm diff})$. Assuming, for simplicity, that the redshifted temperature is uniform (although this is generally not the case
\footnote{The thermal conductivity timescale is larger than the inspiral timescale, so the temperature profile should evolve with time depending on the profile of the energy dissipation rate and that of the heat capacity. Our assumption about uniform redshifted temperature leads to some overestimating of the effect of the diffusion, since diffusion is less efficient at high temperatures. 
})
, we find that it evolves in time according to 
\begin{gather}
\label{heatEq}
C\frac{dT^\infty}{dt}=-\dot{E}_{\rm diff},
\end{gather}
where $C$ is the heat capacity of the star. 

Another important effect is that diffusion leads to the decrease of the orbital energy $E_{\rm orb}$ of the binary system. Here by the orbital energy we imply the sum of the kinetic energies of the star and its companion revolving around their common center of mass, as well as the energy of their gravitational interaction (see Appendix \ref{AppC} for the precise definition of the orbital energy and general evolution equations of the whole binary system in Newtonian framework). For simplicity, in this Section we 
consider the companion as a point-mass object. The total energy, $E_{\rm tot}$, should be conserved, $\dot{E}_{\rm tot}=0$. Within made approximations, this means that the change of the orbital energy $\dot{E}_{\rm orb}$ and mechanical energy of the neutron star, $\dot{E}_{\rm mech}$, is balanced by the heat generation $(-\dot{E}_{\rm diff})$ and energy emitted in the form of gravitational waves $\dot{E}_{\rm GW}$:
\begin{gather}
\dot{E}_{\rm tot}=\dot{E}_{\rm orb}+\dot{E}_{\rm mech}-\dot{E}_{\rm diff}+\dot{E}_{\rm GW}=0.
\end{gather}
Solving this equation for $\dot{E}_{\rm orb}$ and averaging over the orbital period $P_{\rm orb}$, we find
\begin{gather}
\langle\dot{E}_{\rm orb}\rangle=-\langle\dot{E}_{\rm GW}\rangle-\langle\dot{E}_{\rm mech}\rangle+\langle\dot{E}_{\rm diff}\rangle, \qquad \langle f\rangle\equiv\frac{1}{P_{\rm orb}}\int\limits_0^{P_{\rm orb}} f dt.
\end{gather}

Gravitational radiation energy $\dot{E}_{\rm GW}$, strictly speaking, differs from that in the  absence of diffusion. The corresponding corrections to $\dot{E}_{\rm GW}$ due to diffusion are related to the time delay in fluid perturbations caused by dissipation, resulting in a bulge inclination with respect to the binary axis (see, e.g., Refs.\ \cite{gp68,zahn08} and Appendix \ref{AppD}). One can show, however, that these corrections are negligible compared to $\dot{E}_{\rm diff}$ (according to our estimates, they are of the second order in diffusion coefficients).

Next, diffusion influences the mechanical energy of the star by modifying ${\bs\xi}$ and bringing additional contribution to the mechanical energy, see the third term in Eq.\ \eqref{Emech}. However, the corresponding change in $\langle\dot{E}_{\rm mech}\rangle$ does not exceed $\langle{E}_{\rm mech}\rangle/\tau_{\rm evol}$, where $\tau_{\rm evol}$ is the typical time of binary evolution. At the same time, $\langle\dot{E}_{\rm diff}\rangle$ can be estimated as $\langle{E}_{\rm mech}\rangle/\tau_{\rm diff}$, 
where $\tau_{\rm diff}$ is the typical timescale 
on which mechanical energy $E_{\rm mech}$ is converted into heat due to the diffusion process [see Eq.\ (\ref{tauDiff}) below].
As we 
find in our numerical calculations
$\tau_{\rm diff}\ll \tau_{\rm evol}$, which allows one to skip the contribution of diffusion to $\langle\dot{E}_{\rm mech}\rangle$ in comparison to $\langle\dot{E}_{\rm diff}\rangle$.%
\footnote{This statement should not confuse the reader. Despite the fact that $E_{\rm mech}$ dissipates due to diffusion and, seemingly, should decrease on a timescale $\tau_{\rm diff}$, this does not happen due to the continuous replenishment of $E_{\rm mech}$ at the expense of the orbital energy $E_{\rm orb}$ [the second term in Eq.\ (\ref{Ebalance}) is responsible for this].}
As a result, within made approximations, diffusion affects the orbital energy only through the term $\langle\dot{E}_{\rm diff}\rangle$, and the orbital energy loss rate equals
\begin{gather}
\label{AllDeltaphi}
\langle \dot{E}_{\rm orb} \rangle \approx \langle \dot{E}_{\rm orb, ad}\rangle+\langle \dot{E}_{\rm diff}\rangle,
\end{gather}
where $ \dot{E}_{\rm orb, ad}$ is the total orbital energy loss rate in the absence of diffusion (including the loss rate due to gravitational waves).

The additional decrease of the orbital energy caused by diffusion slowly accelerates the inspiral and, therefore, leaves the imprint on the gravitational-wave signal, emitted by the binary. Specifically, it leads to a constantly growing shift in the phase $\phi$ of the gravitational wave signal compared to that in the non-dissipative case. To estimate this effect, let us assume that at some initial moment of time the orbit of the binary is circular, and let us introduce the orbital angular velocity of the binary, $\Omega=2\pi/P_{\rm orb}$. In this case, during the evolution of the binary the orbit will remain circular (see, e.g., Ref.\ \cite{zahn08}) and we can consider all the energy change rates, discussed above, as functions of $\Omega$. Keeping this in mind and using that the gravitational signal frequency equals $2\pi\nu_{\rm GW}=\dot{\phi}=2\Omega(t)$, the phase of the gravitational-wave signal, accumulated while its frequency evolves from $2\pi\nu_{\rm GW, 1}=2\Omega_1$ to $2\pi\nu_{\rm GW,2}=2\Omega_2$, can be found as (in what follows we, for brevity, omit the averaging $\langle\dots\rangle$)
\begin{gather}
\label{delta phi}
\Delta\phi=\int \limits_{\Omega_1}^{\Omega_2} 2\Omega \frac{d\Omega}{\dot{\Omega}}=\int\limits_{\Omega_1}^{\Omega_2} \frac{2\Omega d\Omega}{(d\Omega/dE_{\rm orb})\dot{E}_{\rm orb}}\approx 
\int\limits_{\Omega_1}^{\Omega_2} \frac{2\Omega d\Omega}{(d\Omega/dE_{\rm orb})\dot{E}_{\rm orb, ad}}\biggl[1-\frac{\dot{E}_{\rm diff}}{\dot{E}_{\rm orb, ad}}\biggr].
\end{gather}
Therefore, the contribution of the diffusion to the phase shift can be approximately written as
\begin{gather}
\label{phaseShift}
\Delta \phi_{\rm diff}\approx -\int\limits_{t_1}^{t_2} 2\Omega \frac{\dot E_{\rm diff}}{\dot E_{\rm orb}} \, dt,
\end{gather}
where  $t_1$ and $t_2$ correspond to the moments of time at which the orbital frequencies are $\Omega_1$ and $\Omega_2$, respectively. These moments, as well as the functions $\Omega(t)$ and $\dot E_{\rm orb}$, can be calculated in the approximation of point masses. 

Apart from diffusion, other mechanisms may also contribute to the accumulated phase $\Delta\phi$. Among the dissipative ones diffusion, however, seems to be the leading contributor to the phase shift, at least at large orbit separations, when the orbital angular velocity is low. Indeed, note that dissipation due to diffusion does not depend on the perturbation frequency. At the same time, the dissipation due to shear viscosity is proportional to the perturbation frequency squared, since the former quadratically depends on the velocity gradients. Similarly, dissipation due to mutual friction
\footnote{The dissipative mechanism arising due to 
particle scattering off the neutron vortices in rotating superfluid NSs.}
depends on the relative velocities of vortices and nonsuperfluid particles and is also proportional to the  perturbation frequency squared. Since at large orbit separations the frequency of the oscillations, driven by the tidal forces, is much smaller than the eigenfrequencies of free stellar oscillations, we expect that contributions of shear viscosity and mutual friction to the phase shift are negligibly small compared to $\dot{E}_{\rm diff}$. One can also safely ignore bulk viscosity, since beta-processes are strongly suppressed by proton superconductivity. There are many other effects apart from the dissipative ones, which also contribute to the phase shift
(see, e.g., Ref.\ \cite{Dietrich21}). Here, however, we ignore them and limit ourselves to the phase shift caused by diffusion.


\subsection{Calculating $\dot{E}_{\rm diff}$}\label{sec24}

Thus, in order to determine the phase shift $\Delta\phi_{\rm diff}$, we need first to find the Lagrangian displacement ${\bs\xi}$ and then insert it into the expression for $\dot{E}_{\rm diff}$. 
As long as we deal with small dissipative effects (see discussion below in Sec.\, \ref{validity}) we can adopt a perturbative approach.
Namely, since $\dot{E}_{\rm diff}$ is already linear in diffusion coefficients $D_{km}$,
we can use the Lagrangian displacement of nondissipative theory (accounting for diffusion in $\bs\xi$ would lead to the terms of higher order in $D_{km}$). Thus, our aim in this section is to find the solution of the nondissipative Euler equation
\begin{gather}
\label{EulerEqNoDiff}
e^{-\nu_0/2}(P_0+\varepsilon_0)\ddot{\bs\xi}-\hat{\mathcal{L}}{\bs\xi}=-e^{\nu_0/2}(P_0+\varepsilon_0){\pmb\nabla}\tilde{U},
\end{gather}
satisfying the previously discussed boundary conditions. As mentioned above, we consider the companion star as a point-mass object. Let us choose the orientation of the coordinate system in such a way that $\theta=\pi/2$ describes the equatorial plane of the binary orbit (recall that the origin of the coordinate system is placed in the center of mass of the neutron star). Then in the $(r,\theta,\varphi)$-coordinates the radius vector of the companion star is ${\bs D}(t)=\{D(t),\pi/2,\Phi(t)\}$, and the  gravitational potential, introduced by the companion star inside the considered neutron star, equals
\begin{gather}
U=-\frac{GM'}{|{\bs r}-{\bs D}(t)|}=-GM'\sum_{lm}W_{lm}\frac{r^l}{D(t)^{l+1}}{e}^{-i m \Phi(t)}Y_{lm}(\theta,\varphi), \label{Uexp}
\end{gather}
where $G$ is the gravitational constant, $M'$ is the mass of the companion, $Y_{lm}$ are the spherical harmonics and $W_{lm}$ are some known numerical coefficients (see, e.g., Refs.\ \cite{Lai94}). The $l=m=0$ term in this expansion does not depend on $\bs r$ and thus does not produce any acceleration. The sum of all the $l=1$ terms equals $U_{l=1}=-(G M'/D^3)({\bs D}\cdot{\bs r})$ and results in the uniform acceleration. Therefore, by definition, $l=0$ and $l=1$ terms should be attributed to $U_0$ in the decomposition $U=U_0+\tilde{U}$ [recall the discussion after Eq.\ \eqref{gtensor}], and the tidal potential is given by
\begin{gather}
\tilde{U}=-GM'\sum_{l\geq 2, m}W_{lm}\frac{r^l}{D(t)^{l+1}}{e}^{-i m \Phi(t)}Y_{lm}(\theta,\varphi). \label{Utilde}
\end{gather}

In order to calculate ${\bs \xi}$, we closely follow Ref.\ \cite{Lai94} (see also Refs.\ \cite{pt77,rg94}). As a first step, we expand the Lagrangian displacement into the basis $\{{\bs\xi}_{\bf k}\}$ of complex eigenfunctions ${\bs\xi}_{\bf k}$ of the operator $\hat{\mathcal{L}}$, where the index $\bf k$ labels different eigenfunctions:
\begin{gather}
\label{basis}
{\bs\xi}(t,{\bs r})=\sum_{\bf k}a_{\bf k}(t){\bs \xi}_{\bf k}({\bs r}), \qquad
\left\{
\begin{gathered}
\hat{\mathcal{L}}{\bs\xi}_{\bf k}+e^{-\nu_0/2}(P_0+\varepsilon_0)\omega_{\bf k}^2{\bs\xi}_{\bf k}=0, \\
\gamma P_0 \divv{\bs\xi}_{\bf k}|_{r=R}=0, \\
{\bs\xi}_{\bf k} \text{ is regular at $r=0$}.
\end{gathered}
\right.
\end{gather}
It is easy to see that the numbers $\omega_{\bf k}$ represent the eigenfrequencies of the freely oscillating neutron star, and ${\bs\xi}_{\bf k}$ describe the coordinate dependence of the corresponding eigenfunctions (note that, in such basis, the boundary conditions for ${\bs\xi}$ are satisfied automatically). The eigenmodes of a free neutron star are well known and can be labelled by ${\bf k}=\{nlm\}$, where $n$ is the radial quantum number, while the quantum numbers $l$ and $m$ describe its angular dependence: ${\bs\xi}_{\bf k}={\bs\xi}_{nlm}=[\xi^r_{nl}(r){\bs r}/r+\xi^\perp_{nl}(r)r{\pmb \nabla}]Y_{lm}$.

It is straightforward to verify that the symmetry of the operator $\hat{\mathcal{L}}$ leads to the orthogonality of the eigenmodes with different $\bf k$. In what follows, we imply that the eigenmodes are normalized in such a way that the basis is orthonormal:
\begin{gather}
\label{orthonorm}
\int e^{-\nu_0/2}(P_0+\varepsilon_0){\bs\xi}^\star_{\bf k}{\bs\xi}_{\bf k'} dV=\delta_{\bf k k'}.
\end{gather}
Expanding $\bs\xi$ into the introduced basis, multiplying the Euler equation \eqref{EulerEqNoDiff} by ${\bs\xi}_{\bf k}^\star$, integrating over the stellar volume and using the basis orthonormality, we arrive at the following equation for the coefficient $a_{\bf k}$:
\begin{gather}
\label{EulerEqak}
\ddot a_{\bf k} + \omega_{\bf k}^2 a_{\bf k}=\frac{GM'W_{\bf k}Q_{\bf k}}{D(t)^{l+1}}{e}^{-i m \Phi(t)}, \qquad Q_{\bf k}=\int dV {e}^{\nu_0/2}(P_0+\varepsilon_0) \pmb \xi_{\bf k}^\star \pmb \nabla[r^l Y_{\bf k}(\theta,\varphi)].
\end{gather}
One can show that overlap integrals $Q_{\bf k}$ take real values and do not depend on $m$. As we shall see, they characterize the efficiency of the ${\bf k}$-mode excitation by the tidal potential. Here we would like to note that the orthonormality condition \eqref{orthonorm} does not fix the sign of the amplitude of ${\bs\xi}_{\bf k}$. In what follows, for a given $\bf k$, we eliminate this ambiguity by demanding that $Q_{\bf k}>0$.

To proceed further, we, following Ref.\ \cite{Lai94}, introduce the function $b_{\bf k}(t)$ by
\begin{gather}
\label{bkint}
a_{\bf k}(t)=GM'W_{\bf k}Q_{\bf k} b_{\bf k}(t)e^{-i m \Phi(t)}. 
\end{gather}
Substituting this decomposition into \eqref{EulerEqak}, we obtain
\begin{gather}
\label{EulerEqbk}
\ddot b_{\bf k} -2 i m \Omega \dot b_{\bf k}+(\omega_{\bf k}^2-m^2\Omega^2-i m \dot\Omega) b_{\bf k}=\frac{1}{D(t)^{l+1}},
\end{gather}
where we have used that $\dot{\Phi}=\Omega$. Strictly speaking, we should integrate this equation numerically. In the limit of large orbit separations, while $\omega_{\bf k}>m\Omega$, it is, however, safe to use the asymptotic solution instead. To obtain this solution, we note that, due to slow evolution of the binary, all the terms containing $d/dt$ can be discarded as small. As a result, we find
\begin{gather}
\label{bkAsympt}
b_{\bf k}=\frac{1}{D^{l+1}(\omega_{\bf k}^2-m^2\Omega^2)}. 
\end{gather}
Thus, equations \eqref{bkAsympt}, \eqref{bkint} and \eqref{basis} define the Lagrangian displacement in the limit of large orbit separations. Further, we use the obtained ${\bs\xi}$ to calculate the dissipation rate $\dot{E}_{\rm diff}$. Using that $\bs\xi$ is real-valued (although the basis eigenfunctions are complex), we have
\begin{gather}
\label{Ediffak}
\dot{E}_{\rm diff}=\int {\bs\xi}\hat{\mathcal{D}}{\bs\xi} dV=\int {\bs\xi}^\star\hat{\mathcal{D}}{\bs\xi} dV=\sum_{\bf k \bf k'}a_{\bf k}(t)a_{\bf k'}^\star(t)\int {\bs\xi}^\star_{\bf k'}\hat{\mathcal{D}}{\bs\xi}_{\bf k} dV.
\end{gather}
Although the cross terms of modes with different angular dependence (i.e., quantum numbers $l$ and $m$) vanish, the cross-terms of different modes with the same angular dependence but different radial numbers $n$, generally, do not vanish. In the case of free neutron star oscillations $a_{\bf k}\propto e^{i\omega_{\bf k}t}$ and they would vanish after averaging over the large enough period of time $\Delta t\gg |\omega_{\bf k}-\omega_{\bf k'}|^{-1}$. Here we consider oscillations driven by the tidal force at the early stages of the binary inspiral. In this case, non-vanishing cross-terms have $a_{\bf k}\propto a_{\bf k'}\propto e^{i m \Omega t}$ and, generally, do not vanish even after time-averaging. 

All the equations, discussed above, are written for the binary system with a point-mass companion. Some of the obtained results, however, can easily be generalized to the case of a binary system with two completely identical neutron stars. Clearly, the total heat loss in such a binary is twice larger than that in a binary with a point-mass companion. 
The same applies also to the total phase shift $\Delta\phi_{\rm diff}$ of the gravitational signal due to diffusion.



\section{Numerical results}\label{results}

In this section, we provide the numerical results obtained for the phase shift $\Delta\phi_{\rm diff}$ of the gravitational signal and investigate the signal's potential detectability by the Einstein Telescope. In our numerical calculations, we consider a binary system of two $M=1.0 \, M_\odot$ or two $M=1.4 \, M_\odot$ identical strongly superconducting neutron stars, each described by the BSk24 equation of state \cite{gcp13,pcp18}. We employ diffusion coefficients $D_{km}$ from Ref.\ \cite{dgs20}, assuming that strong proton superconductivity effectively reduces the number of particle species by one \cite{gg23}. That is, the coefficients $D_{km}$ in superconducting four-component $npe\mu$-matter are calculated by the formulas of Ref.\ \cite{dgs20} for the three-component $ne\mu$-matter
\footnote{We also neglect the effects of proton superconductivity and Fermi-liquid effects on the momentum transfer rate between neutrons and electrons, $J_{ne}$, neutrons and muons, $J_{n\mu}$, and muons and electrons, $J_{e\mu}$. }.
To calculate the heat capacity $C$ of a neutron star, we follow Ref.\ \cite{yls99}.


\subsection{Phase shift of GW signal and NS heating due to diffusion}

In order to determine the phase shift $\Delta\phi_{\rm diff}$, we should find the basis functions $\{{\bs\xi}_{\bf k}\}$ and eigenfrequencies $\{\omega_{\bf k}\}$ by solving the eigenvalue problem \eqref{basis}. In the Cowling classification \cite{cowling41} of stellar oscillations the corresponding eigenfunctions are known as $f$-modes, $p$-modes, and $g$-modes. According to \eqref{basis}, the tidal potential perturbs modes with the same spatial dependence as in an isolated neutron star but with a different time dependence. In what follows, these modes are also referred to as $f$-, $p$- and $g$-modes despite their differences from the free neutron star eigenmodes. The efficiency of the excitation of one mode or another by the tidal potential depends on the values of the corresponding overlap integrals $Q_{\bf k}$ and eigenfrequencies $\omega_{\bf k}$ [see Eqs.\ \eqref{EulerEqak}, \eqref{bkint}, and \eqref{bkAsympt}]. According to our numerical calculations, the $\{l,m\}=\{2,0\}$ and $\{l,m\}=\{2,\pm 2\}$ $f$-modes, and $g$-modes are the main contributors to the tidal perturbation ${\bs\xi}$ \eqref{basis}. Some values of the corresponding overlap integrals, $Q_{\bf k}$, as well as the ratios $Q_{\bf k}/\omega_{\bf k}^2$ (which drive the amplitude of the perturbation, see Eqs.\ \ref{bkint} and \ref{bkAsympt}) are given in Table \ref{overlapVals}. Here we would like to note that contributions with $l>2$ can be ignored as negligibly small, since they are of higher order in $R/D$ [see Eq.\ \eqref{bkAsympt}]. The $\{l,m\}=\{2,\pm 1\}$ terms also do not contribute, since $W_{2,\pm 1}=0$ [see Eq.\ \eqref{EulerEqak}]. Finally, $p$-modes possess eigenfrequencies much higher than those of $g$-modes and, as a result, are excited less efficiently and also can be ignored.


\begin{table}
\centering
\begin{tabular}{|c|c|c|c|c|}
\multicolumn{5}{c}{ $l=2$ overlap integrals} \\ \hline
Mass         & \multicolumn{2}{|c|}{$1.4 \, M_\odot$}              & \multicolumn{2}{|c|}{$1.0 \, M_\odot$}              \\ \hline
Quantity     & $Q_{\bf k}$         & $Q_{\bf k}/\omega_{\bf k}^2$  & $Q_{\bf k}$          & $Q_{\bf k}/\omega_{\bf k}^2$ \\ \hline
$f$-mode     & $4.0\times 10^{-1}$ & $2.0\times 10^{-1}$           & $4.4 \times 10^{-1}$ & $1.8 \times 10^{-1}$ \\ \hline
$g_0$-mode   & $2.4\times 10^{-4}$ & $1.8\times 10^{-2}$           & $3.6 \times 10^{-4}$ & $2.7 \times 10^{-2}$ \\ \hline
$g_1$-mode   & $1.3\times 10^{-4}$ & $1.9\times 10^{-2}$           & $1.3 \times 10^{-4}$ & $2.1 \times 10^{-2}$ \\ \hline
$g_2$-mode   & $6.7\times 10^{-5}$ & $1.8\times 10^{-2}$           & $6.7 \times 10^{-5}$ & $1.9 \times 10^{-2}$ \\ \hline
$g_3$-mode   & $4.3\times 10^{-5}$ & $1.7\times 10^{-2}$           & $4.7 \times 10^{-5}$ & $2.2 \times 10^{-2}$ \\ \hline
$g_4$-mode   & $3.9\times 10^{-5}$ & $2.3\times 10^{-2}$           & $4.2 \times 10^{-5}$ & $2.8 \times 10^{-2}$ \\ \hline
\end{tabular}
\caption{Overlap integrals $Q_{\bf k}$ and ratios $Q_{\bf k}/\omega_{\bf k}^2$ for the $l=2$ $f$-modes and some $l=2$ $g$-modes. The main $g$-mode harmonic is denoted as $g_0$, while the $n$-th $g$-mode overtones are denoted as $g_n$. The eigenfrequencies $\omega_{\bf k}$ are normalized to $\sqrt{GM/R^3}$.}
\label{overlapVals}
\end{table}


As a next step, we need to calculate $\dot{E}_{\rm diff}(t)$ for a given tidal perturbation. Strictly speaking, the fact that $f$-modes are perturbed much more efficiently than the other modes, as follows from Table \ref{overlapVals}, does not necessarily imply that they contribute the most to $\dot{E}_{\rm diff}$. In fact, as shown in Ref.\ \cite{kgk21}, in case of an isolated neutron star $g$-modes and $p$-modes dissipate energy through diffusion much faster than $f$-modes do. The reason is that the chemical potential imbalances are only weakly perturbed for practically incompressible $f$-modes and cannot lead to significant dissipation [see discussion after Eq.\ \eqref{EdiffMu}]. The same is true for the $f$-modes perturbed by tidal potential, since they have the same spatial dependence. As a result, although $f$-modes are the main contributors to the tidal perturbation, the $g$-modes could, in principle, dissipate strongly and their contribution to the dissipation could be of the same order as that of $f$-modes. Nevertheless, in our calculations of $\dot{E}_{\rm diff}$, we retain only the $m=\pm 2$ $f$-mode contributions to the tidal perturbation. As we verified numerically (see below), this approximation is justified and allows us to obtain a reasonable estimate for $\Delta\phi_{\rm diff}$.


\begin{figure}
\center{\includegraphics[width=0.7\linewidth]{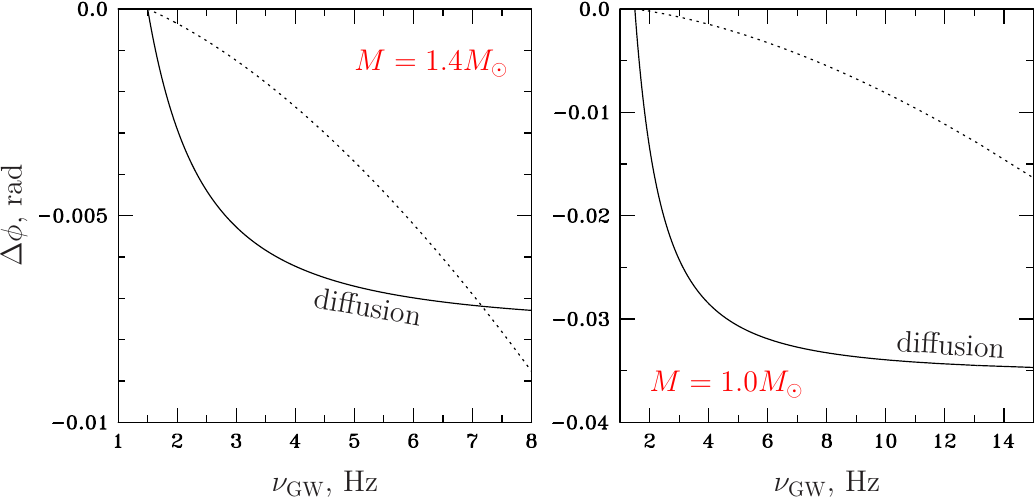}}
\caption{Solid and dashed lines show the phase shifts due to diffusion and due to gravitational radiation of the NS bulge, respectively, as functions of GW frequency. Two panels correspond to (from left to right) $M=1.4M_\odot$ and $M=1.0M_\odot$.}
\label{DeltaphiFig}
\end{figure}


In order to find $\dot{E}_{\rm diff}(t)$ we note that, at any moment of time, $\dot E_{\rm diff}\propto (T^\infty)^{-2}$ with high accuracy. This dependence reveals itself in numerical calculations with $D_{km}$ taken from Ref.\ \cite{dgs20} (see also Appendix \ref{AppE}). It allows one to easily find $\dot{E}_{\rm diff}(t)$ by integrating Eq.\ \eqref{heatEq}. Once the initial value of $T^\infty$ is given, one can calculate the corresponding initial value of $\dot{E}_{\rm diff}$ from Eq.\ \eqref{EdiffMu} and establish the proportionality coefficient in $\dot E_{\rm diff}\propto (T^\infty)^{-2}$. After that the integration of \eqref{heatEq} becomes trivial. 
In our calculations, as an initial value at sufficiently large separations 
\footnote{
We start the integration at the separation corresponding to $\nu_{\rm GW}\approx 0.1\,\rm Hz$. The results for temperature are not sensitive to the choice of the starting value of $\nu_{\rm GW}$, as long as it is sufficiently low.}%
we adopt
$T^\infty=10^6\,\rm K$, which, depending on the envelope composition, corresponds to the effective redshifted surface temperature $(0.82-0.89)\times 10^5\,\rm K$ in case of $M=1.4M_\odot$ and $(0.78-0.83)\times 10^5\,\rm K$ in case of $M=1.0M_\odot$, respectively \cite{pcy97}. Once $\dot{E}_{\rm diff}$ is found, we calculate the phase shift $\Delta\phi_{\rm diff}$ from Eq.\ \eqref{phaseShift} (recall that the result should be multiplied by $2$, since we consider a binary of two identitcal neutron stars, and not the binary with a point-mass companion). The resulting phase shifts for the $M=1.0 \, M_\odot$ and $M=1.4 \, M_\odot$ neutron star models are shown as functions of the gravitational-wave signal frequency $\nu_{\rm GW}$ in Fig.\ \ref{DeltaphiFig} by solid lines. Here it is assumed that the observation starts at $\nu_{\rm GW}=1.5\,\rm Hz$
\footnote{Our choice of $\nu_{\rm GW}=1.5\,\rm Hz$ as the starting frequency of the observation was guided by the sensitivity curves of the Einstein telescope \cite{et12,telescopes19a,telescopes19b}.}. In the same figure, dashed lines, for comparison, show the contribution of the gravitational radiation of the NS bulge calculated with equation (51) of Ref.\ \cite{Dietrich21}. To plot the dashed curves, we adopted the tidal deformability from Ref.\ \cite{pc21}. Notably, while at $\nu_{\rm GW}\sim$ hundreds of hertz this is the main contribution to the phase shift due to matter effects, at low frequencies considered here the contribution of dissipation is more important, especially for light NSs. For $M=1.0M_\odot$ NS model $\Delta \phi_{\rm diff}$ reaches a few hundredths of a radian. 
 
\begin{figure}
\centering
\includegraphics[width=0.7\linewidth]{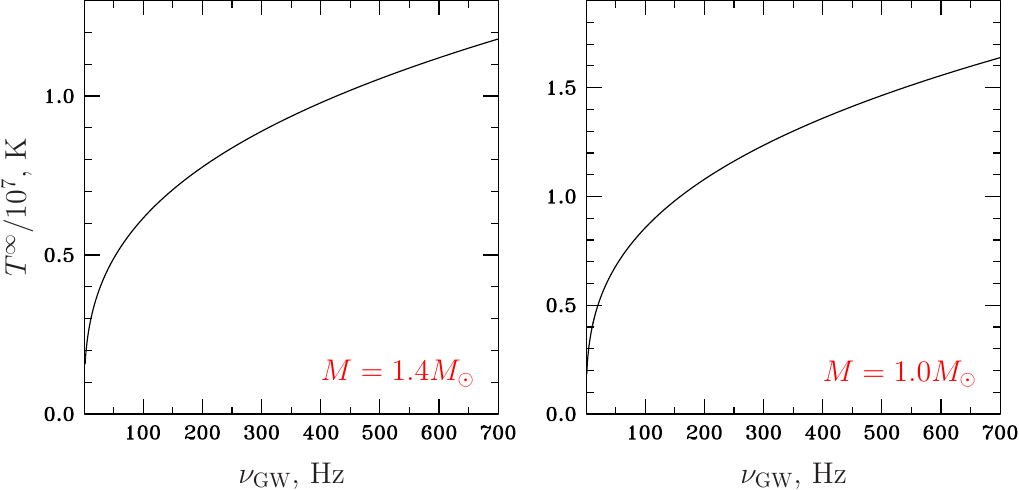}
\caption{Internal stellar temperature versus the frequency of the emitted gravitational wave for the same two stellar models as in Fig.\ \ref{DeltaphiFig}.}
\label{Tpic}
\end{figure}

Further, using the calculated $\dot{E}_{\rm diff}$, one can also estimate the NS heating due to diffusion. Figure \ref{Tpic} shows $T^\infty$ versus $\nu_{\rm GW}$ for the two considered stellar models. Our results imply that particle diffusion may heat NS up to values $\sim 10^7\,\rm K$. Previous calculations of the heating due to viscous dissipation of $f$- and resonantly excited $g$-modes resulted in a similar estimate for the temperature, see, e.g., Refs.\ \cite{yw17,kksk22}.


\subsection{Approximate formula for the phase shift and GW signal detectability at low frequencies}

\begin{figure}
\centering
\includegraphics[width=0.35\linewidth]{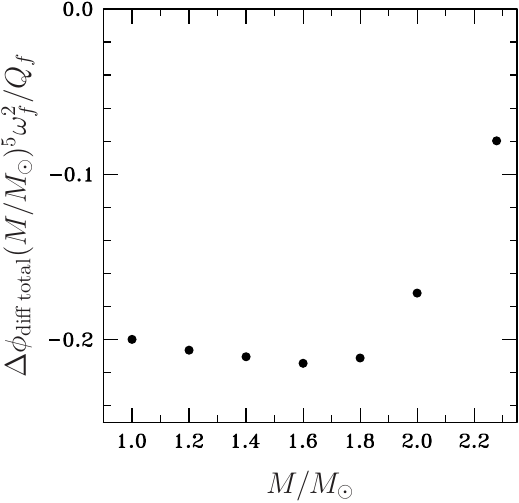}
\caption{$\Delta \phi_{\rm diff\,total} (M/M_\odot)^{5}\omega_{f}^{2}/Q_{f}$ as a function of $M/M_\odot$. Here $\omega_{f}$ is normalized to $\sqrt{GM/R^3}$.}
\label{App}
\end{figure}

Let us consider the phase shift $\Delta\phi_{\rm diff\, total}$ accumulated during the whole inspiral, and estimate its dependence on the stellar parameters. For simplicity, we assume that inspiralling stars have equal masses, $M$.
As a first step, we change the integration variable in \eqref{phaseShift}  from $t$ to $\Omega$: $dt=d\Omega/\dot{\Omega}$, where $\dot{\Omega}$ is calculated using the point mass approximation. Next, to find $\dot{E}_{\rm orb}$, we also use the point mass approximation, where $E_{\rm orb}=-GM^2/(2D)$: $\dot{E}_{\rm orb}=(dE_{\rm orb}/dD)\dot{D}$. Explicit expressions for $\dot{\Omega}$ and $\dot{D}$ in the point-mass approximation can be found, e.g., in Ref.\ \cite{Lai94}. Finally, we substitute the expression \eqref{Ediffak} into the integrand, accounting only for the $l=2$ $f$-mode contributions. Expressing $a_{\bf k}$ with Eqs.\ \eqref{bkint} and \eqref{bkAsympt}, and using the fact that $\Omega^2\propto M/D^3$, we eventually find
\begin{gather}
\Delta \phi_{\rm diff\,total}\propto\int\limits_{\Omega_1}^{\Omega_2} \frac{Q_{f}^2 }{\omega_{f}^4 M^{5}}\frac{1}{\Omega^2 (T^\infty)^{2}}\left(\int {\bs\xi}_{f}^\star (T^\infty)^{2}\hat{\mathcal{D}}{\bs\xi}_{f} dV\right) d\Omega, \label{approx1}
\end{gather}
where the index $f$ refers to the $l=m=2$ $f$-mode, while $\Omega_1$ and $\Omega_2$ are the orbital frequencies in the beginning and in the end of the observation (in our calculations $\Omega_1/\pi=1.5\, \rm Hz$). 
The operator $\hat{\mathcal{D}}$ is $\propto (T^\infty)^{-2}$, thus to make the integral temperature-independent, we factored out its temperature dependence. To find how the temperature depends on $\Omega$ we consider the evolution laws of the corresponding quantities: 
\begin{gather}
    \dot \Omega \propto \Omega^{11/3} M^{5/3}, \label{Omegaprop}\\
    \dot T^\infty\propto \frac{Q_{f}^2}{\omega_{f}^4}\Omega^4\frac{1}{(C/T^\infty) \, (T^\infty)^3}\int {\bs\xi}_{f}^\star (T^\infty)^2\hat{\mathcal{D}}{\bs\xi}_{f} dV. \label{Tprop}
\end{gather} 
Here $C$ is the total heat capacity of the star, the ratio $C/T^\infty$ is temperature-independent, and, as we checked numerically, $C\propto M$. 
Equation (\ref{Omegaprop}) describes the orbital frequency evolution in the point-mass approximation, see, e.g., Ref.\ \cite{Lai94}.
Equation (\ref{Tprop}) is a consequence of the thermal balance equation (\ref{heatEq}) with Lagrangian displacements entering $\dot E_{\rm diss}$ given by Eq.\ (\ref{basis}), (\ref{bkint}), and (\ref{bkAsympt}).
Integrating the ratio of (\ref{Omegaprop}) and (\ref{Tprop}), we find
\begin{gather}
    (T^\infty)^2 \propto \Omega^{2/3} M^{-4/3}\frac{Q_{f}}{\omega_{f}^2}\left(-\int {\bs\xi}_{f}^\star (T^\infty)^2\hat{\mathcal{D}}{\bs\xi}_{f} dV\right)^{1/2}.
\end{gather} 
Introducing this dependence into (\ref{approx1}) and performing integration we arrive at (assuming $\Omega_2\gg \Omega_1$):
\begin{gather}
\Delta \phi_{\rm diff\,total}
\propto
\frac{Q_{f}}{\omega_{f}^2 M^{11/3}}\frac{1}{\Omega_1^{5/3}}\left(-\int {\bs\xi}_{f}^\star (T^\infty)^2\hat{\mathcal{D}}{\bs\xi}_{f} dV\right)^{1/2}.   \label{approx}
\end{gather}
While the dependence of $\Delta \phi_{\rm diff\,total}$ on  $\Omega_1$ is 
well-established,
its dependence on stellar
parameters is less certain, since the integral in Eq.\ (\ref{approx}) may 
vary with the
stellar mass and it is not so trivial to parameterize this dependence. 
That is why we fit our numerical results with a simple formula
\begin{gather}
\label{phiApprox}
\Delta \phi_{\rm diff\,total}\approx -0.5\frac{1}{\Omega_1^{5/3}}\frac{Q_{f}/\omega_{f}^{2}}{0.2} (M/M_\odot)^{-5},
\end{gather}
where $\omega_{f}$ is normalized to $\sqrt{GM/R^3}$ and the ratios $Q_{f}/\omega_{f}^{2}$ (for $M=1.0 M_\odot$ and $M=1.4 M_\odot$) are presented in Table \ref{overlapVals}. 
To illustrate the proportionality law \eqref{phiApprox}, we plot $\Delta \phi_{\rm diff\,total} (M/M_\odot)^{5}\omega_{f}^{2}/Q_{f}$ for various stellar masses in Fig.\ \ref{App}. From this figure we see that for $M<2M_\odot$ the approximation (\ref{phiApprox}) works well for the considered BSk24 EOS.
At higher masses, however, the approximation \eqref{phiApprox} is less accurate. Note also that, depending on the EOS, $\Delta \phi_{\rm diff\,total}$ may deviate from this simple formula, exhibiting a slightly different dependence on stellar mass and an uncertainty of a factor of a few in the numerical multiplier. The difference is caused by the integral in \eqref{approx} which generally depends on the stellar model parameters in a nontrivial way.


\begin{figure}
\centering
\includegraphics[width=0.35\linewidth]{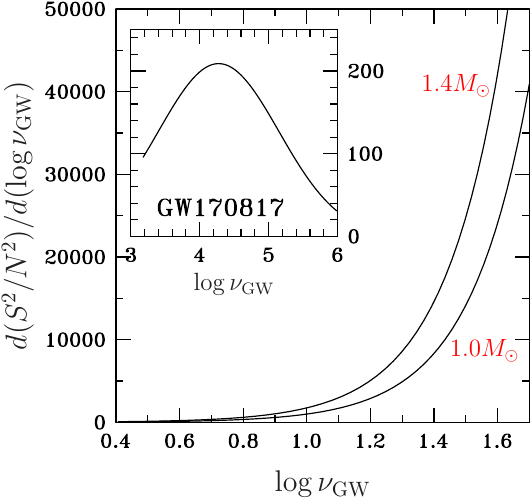}
\caption{Distribution of signal-to-noise squared defined by Eq.\ (\ref{SNnew}) in ${\rm log}\,\nu_{\rm GW}$ for the Einstein Telescope. The curves are calculated for a binary inspiral at $d=40 \,\rm Mpc$. Two curves correspond to two stellar models ($M=1.4M_\odot$ and $M=1.0M_\odot$). The inset shows the same distribution for GW170817 event and advanced LIGO parameters ($M=1.4M_\odot$ is adopted for this curve).}
\label{SignalToNoise}
\end{figure}

Before concluding this section, let us estimate whether the gravitational signal from the binary can be detected by the Einstein Telescope at the low frequencies of interest. To this aim, we analyze the signal-to-noise ratio $S/N$, defined by the formula \cite{mcb15}
\begin{gather}
\frac{S^2}{N^2}=\int\limits_{0}^\infty\frac{2\nu_{\rm GW}\,h_0^2}{\dot \nu_{\rm GW}\,S_{\rm n}(\nu_{\rm GW})}\frac{d \nu_{\rm GW}}{\nu_{\rm GW}}= \int\limits_{\nu_{\rm GW}=0}^{\nu_{\rm GW}=\infty}\frac{2\nu_{\rm GW}\,h_0^2}{\dot \nu_{\rm GW}\,S_{\rm n}(\nu_{\rm GW})}  d({\rm log}\, \nu_{\rm GW}). \label{SNnew}
\end{gather}
Here $S_{\rm n}(\nu_{\rm GW})$ is the power spectral density of the detector noise. In our calculations, we adopt $S_{\rm n}(\nu_{\rm GW})$ for the Einstein Telescope from Refs.\ \cite{et12,telescopes19a,telescopes19b}
\footnote{One should bear in mind that in practice, noise at low frequencies may turn out to be stronger, and, perhaps, our estimates are too optimistic.}. Next, $\nu_{\rm GW}/\dot \nu_{\rm GW}$ is the time during which the binary emits in the bandwidth centered around $\nu_{\rm GW}$. Finally, $h_0$ is the GW amplitude \cite{watts08}
\footnote{Note that $h_0^2$ in Refs.\ \cite{watts08, mcb15} differ by a factor of $2$ from each other.},
%
\begin{gather}
h_0^2=\frac{5G}{4\pi^2 c^3 d^2 \nu_{\rm GW}^2} \dot E_{\rm GW}\approx -\frac{5G}{4\pi^2 c^3 d^2 \nu_{\rm GW}^2} \dot E_{\rm orb},
\end{gather}
where $d$ is the distance to the binary. In Fig.\ \ref{SignalToNoise} we plot the integrand of Eq.\ (\ref{SNnew}), $2\nu_{\rm GW}\,h_0^2/[\dot \nu_{\rm GW}\,S_{\rm n}(\nu_{\rm GW})]$, as a function of ${\rm log}\, \nu_{\rm GW}$. To plot Fig.\ \ref{SignalToNoise} we assume the distance $d$ corresponding to GW170817 event, $d=40\, \rm Mpc$. For the reference, in the inset we also present the same integrand from Eq.\ (\ref{SNnew}) for the Advanced LIGO (assuming that stellar masses equal $1.4M_\odot$ and adopting $S_{\rm n}(\nu_{\rm GW})$ for the Advanced LIGO from Refs.\ \cite{telescopes19a,telescopes19b}). Figure \ref{SignalToNoise} implies that detection of the GW signal from NS inspirals by the Einstein Telescope in the frequency band of a few hertz is very favorable even for inspirals at distances significantly exceeding that of the GW170817 event. This indicates that one can expect numerous detections. This, in turn, gives some hope that the phase shift due to diffusion, $\sim 0.03\,\,\rm rad$, can be detected by statistical methods -- while such a phase shift may be too small to be detected in one observation (see, e.g., Ref.\ \cite{sensitivity18,mzl23,read23,ha23}), it, probably, could be detected in the analysis of a large number of events.

Note, in this respect, that the proposed space-based GW observatory DECIGO \cite{DECIGO} looks even more promising with its greatest sensitivity in the frequency range of $(0.1-10)$~Hz, allowing the phase shift to accumulate over a few years and reach several radians.


\subsection{Validity of our approximations}\label{validity}

In order to check the validity of the approximations made to obtain $\Delta\phi_{\rm diff}$ it is necessary to make sure that the discarded $l=2$ $m=0$ $f$-mode and $l=2$ $g$-modes do not contribute significantly to $\dot{E}_{\rm diff}$.

Let us first discuss the $m=0$ case. For these modes perturbations are very slow and 
evolve
on a timescale of the binary orbit evolution, $\tau_{\rm evol}\sim D/\dot{D}\gg P_{\rm orb}$. At such long timescales diffusive currents of these modes effectively relax matter to diffusion equilibrium.
As a result, $m=0$ modes almost do not drive the neutron star out of diffusion equilibrium and perturbations proceed almost adiabatically, but in the limit opposite to that considered in the paper. According to the discussion after Eq.\ \eqref{EdiffMu}, this implies that diffusive currents $\Delta{\bs j}_k$, induced by these modes, are weak and the contribution of $m=0$ modes to $\dot{E}_{\rm diff}$ is small.

As for the $m=\pm 2$ modes, their perturbation timescale, according to Eq.\ \eqref{EulerEqak}, equals 
$2\pi/(|m|\Omega)=P_{\rm orb}/2$,
which
is much shorter than $\tau_{\rm evol}$. As a result, the diffusive currents calculated within the approach adopted in the paper are indeed small in comparison to adiabatic currents in the whole integration range for $m=\pm 2$ f-modes and first considered harmonics of g-modes (checked numerically). Thus, the equations of Sections \ref{perturb}-\ref{IIC} are valid for these modes. Further, to check the validity of using of the approximate Eq.\ \ref{EulerEqNoDiff} one should ensure that the dissipative term discarded in Eq.\ \ref{EulerEqNoDiff} is indeed small. 
This is equivalent to checking that
$\tau_{\rm diff, \bf k}\gg P_{\rm orb}/2$, where
the diffusive relaxation timescales $\tau_{\rm diff, \bf k}$ 
are defined as
\begin{gather}
\label{tauDiff}
\tau_{\rm diff, \bf k}=-\frac{2 E_{\rm mech \, ad, \bf k}}{\dot{E}_{\rm diff, \bf k}}, \quad
E_{\rm mech \, ad, \bf k}\equiv \frac{1}{2}\dot{a}_{\bf k}^\star\dot{a}_{\bf k}+\frac{1}{2}\omega_{\bf k}^2 a_{\bf k}^\star a_{\bf k} ,\quad
\dot{E}_{\rm diff, \bf k}\equiv \sum_{\bf k'} a_{\bf k}a_{\bf k'}^\star\int {\bs\xi}^\star_{\bf k'}\hat{\mathcal{D}}{\bs\xi}_{\bf k} dV.
\end{gather}
Here $E_{\rm mech \, ad, \bf k}$ is the contribution of the considered mode to the mechanical energy in the non-diffusive limit (obtained by substituting Eq.\ \eqref{basis} into the formula for the mechanical energy (\ref{Emech ad}) and integrating over the stellar volume), and $\dot{E}_{\rm diff, \bf k}$ is the contribution of the mode to the diffusive energy dissipation [cf.\ Eq.\ \eqref{Ediffak}].


\begin{figure}
\centering
\includegraphics[width=0.7\linewidth]{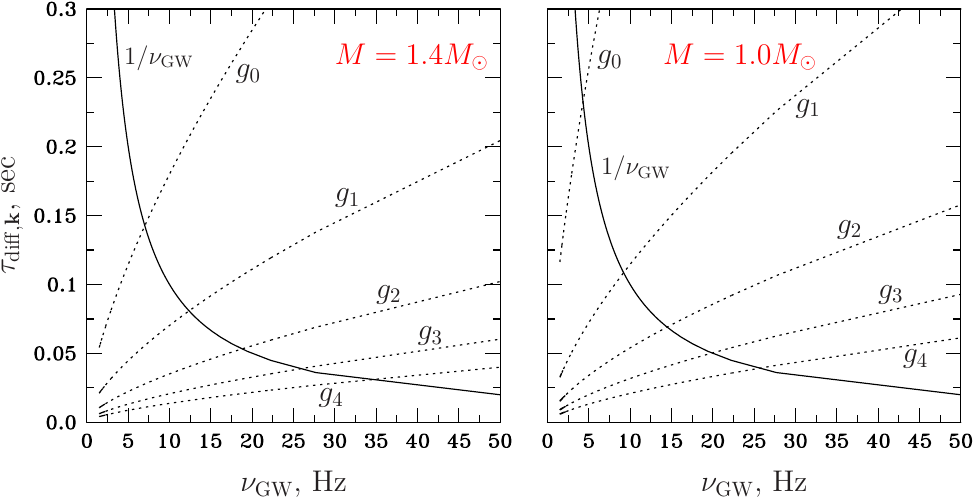}
\caption{Two panels show diffusive relaxation times of various $g$-modes as functions of GW frequency (dashed lines) for two adopted stellar models. Solid lines show the period of $m=\pm 2$ perturbation, $1/\nu_{\rm GW}$.}
\label{timescales}
\end{figure}


To proceed further, we will 
calculate $\tau_{\rm diff, \bf k}$ assuming that Eq.\ \ref{EulerEqNoDiff} is valid for all the modes with $m=\pm 2$ under consideration.
Our numerical calculations indicate that, due to the interplay between the mode amplitudes ($f$-modes have larger amplitudes) and dissipation efficiency ($g$-modes dissipate stronger), the cross-terms in $\dot{E}_{\rm diff, \bf k}$ are of the same order as the term with ${\bf k}={\bf k'}$. Moreover, 
the cross-terms 
may be both positive and negative, 
so that accounting for larger number 
of these terms
does not lead to a monotonic increase of 
$\dot{E}_{\rm diff, \bf k}$; instead, $\dot{E}_{\rm diff, \bf k}$ wanders near the value of the ${\bf k}={\bf k'}$ term. 
In addition,
in our numerical examples, 
$\dot{E}_{\rm diff, \bf k}$ 
is of the same order of magnitude
for all the considered modes.
Finally, let us also point out that one
can safely ignore the kinetic energy of a given mode in comparison to its potential energy, 
since the frequency of forced oscillations of the mode is
much smaller than the corresponding eigenfrequency $\omega_{\bf k}$
of a freely oscillating NS.
For these reasons, one can use the following estimates:
\begin{gather}
E_{\rm mech \, ad, \bf k}\sim \frac{1}{2}\omega_{\bf k}^2 a_{\bf k}^\star a_{\bf k}, \qquad
\dot{E}_{\rm diff, \bf k}\sim a_{\bf k}^\star a_{\bf k}\int {\bs\xi}^\star_{\bf k}\hat{\mathcal{D}}{\bs\xi}_{\bf k} dV, \qquad
\tau_{\rm diff, \bf k}\sim -\frac{\omega_{\bf k}^2}{\int {\bs\xi}^\star_{\bf k}\hat{\mathcal{D}}{\bs\xi}_{\bf k} dV}. \label{tau}
\end{gather}
Note that $\tau_{\rm diff, \bf k}$ depend on $t$ only through the temperature dependence $T^\infty(t)$. The relaxation timescale of the $m=\pm 2$ $f$-modes turns out to be larger than the perturbation timescale and, therefore, the framework can be fully applied. The relaxation timescales of various $g$-modes, obtained using this estimate, are shown by dots in Fig.\ \ref{timescales} as functions of the frequency $\nu_{\rm GW}=\Omega/\pi$ of the emitted gravitational wave 
[$T^\infty(\nu_{\rm GW})$ from Fig.\ \ref{Tpic} was adopted to plot the figure]. 
For reference, solid lines show the perturbation timescale of the $m=\pm 2$ modes, $P_{\rm orb}/2$.
The figure implies that 
for both models at $\nu_{\rm GW}\la 5\,\rm Hz$ all $g$-mode perturbations have $\tau_{\rm diff, \bf k}< P_{\rm orb}/2$.
Strictly speaking this means that Eq.\ \ref{EulerEqNoDiff} is not applicable for g-modes and one has to keep the
dissipative term in the
Euler equation when
calculating 
the mode amplitude (see Appendix \ref{AppD}). Keeping dissipation in Eq.\ \ref{EulerEqNoDiff} would lead to suppression of the mode amplitude in comparison to numerical examples considered in the present paper. This means that the contribution of g-modes to $\dot{E}_{\rm diff}$ would not be of the same order as that of the f-mode (as we observe in our calculations) but substantially lower. 
Thus, we see that at the early stages of the inspiral, where most of the 
diffusion-related
phase shift is accumulated (see Fig.\ \ref{DeltaphiFig}), it is safe to ignore contributions from the $g$-modes and $m=0$ $f$-mode. At later stages, however, $g$-modes come into play one after another, starting with the main harmonic, and should be accounted for in the expression \eqref{EdiffMu}.
However, as we have verified, they do not qualitatively affect $\dot{E}_{\rm diff}$, 
introducing an inaccuracy on the order of a factor of two.
In other words, allowing for only the $\{l, m\}=\{2,\pm 2\}$ $f$-modes provides a reasonable order-of-magnitude estimate
even during the later stages of the inspiral.
In particular, this conclusion applies to the calculation of the stellar temperature (see Fig.\ \ref{Tpic}).


 
\section{Discussion}
\label{disc} 

During NS inspiral, NS matter deviates from equilibrium, being perturbed by the tidal potential. Various dissipative mechanisms tend to relax this perturbation. In this work, we analyzed the role of particle diffusion, which has been shown to be very efficient in superconducting NSs \cite{kgk21,kgk23}. We found that diffusion may affect the phase of the gravitational wave from inspiralling NSs. The effect is especially pronounced if stellar masses are low. 
Ground-based telescopes can detect the phase shift due to diffusion up to a few hundredths of a radian 
(see right panel of Fig.\ \ref{DeltaphiFig} for $M=1.0M_\odot$). Most of the phase shift accumulates at early stages, when the gravitational wave frequency evolves to a few hertz. Currently operating GW observatories are not sensitive in this band, while among the next-generation ground-based observatories, the Einstein Telescope will be the most efficient in this frequency range. 
We verified that it could be able to effectively detect signals from NSs at the early stage of inspiral (see Fig.\ \ref{SignalToNoise}). 
In turn, among the planned {\it space-based} observatories, the most promising one is DECIGO, 
as it will be able to detect the GW signal at earlier stages of inspiral, where the phase shift can reach a few radians.
We also found
that particle diffusion leads to the heating of the stellar matter up to a temperature of $\sim 10^7$~K (see Fig.\ \ref{Tpic}), which is comparable to the findings of other authors \cite{yw17,kksk22} who considered NS heating due to viscous dissipation.

We should warn the reader that these results were obtained assuming that protons are strongly superconducting while neutrons are normal throughout the whole NS core. Superconducting protons do not scatter off neutrons, making high relative velocities between neutrons and charged particles (in particular, leptons) possible. Dissipation occurring due to leptons' scattering off neutrons turns out to be very efficient, thanks to their high relative velocities. However, according to microscopic calculations (see Ref.\ \cite{sc19} and references therein), neutron and proton critical temperatures have density-dependent profiles, and some parts of the core can contain superfluid neutrons, while protons are not necessarily superconducting throughout the entire core. Thus, a detailed analysis accounting for more realistic microphysics is warranted. However, even now, we can make some estimates. At the large orbital separations relevant to our study, the temperature of the neutron star is lower than $10^7\rm K$ (see  Fig.\ \ref{Tpic}). Neutron and proton critical temperatures reach in maximum few$\times 10^8\,\rm K$ and few$\times 10^9\,\rm K$, respectively \cite{drddwcp16,sc19}. This means that some part of the core contains strongly superfluid neutrons. Particle scattering off neutrons is exponentially suppressed there. In $npe$ matter, this leads to the elimination of dissipation caused by diffusion, since protons and electrons are electromagnetically coupled in this case and their relative velocity vanishes with a very high accuracy. The admixture of muons, however, permits relative motions of charged particles, and dissipation is then driven by their scattering off each other. Since momentum transfer rates between charged particles are about two orders of magnitude more efficient than between neutrons and leptons \cite{dgs20}, the particle relative velocities, and hence dissipation, are two orders of magnitude lower \cite{kgk21} than in the case considered in this work. Thus, the core region with strongly superfluid neutrons does not contribute to dissipation effectively. Next, in the region where both neutrons and protons are nonsuperfluid/nonsuperconducting, neutrons are strongly coupled to protons due to their effective scattering by means of the strong interaction, and dissipation occurs again due to scattering of charge particles among themselves (assuming the presence of muons, see above). Consequently, the nonsuperconducting region almost does not contribute to dissipation as well. This means that the phase shift plotted in Fig.\ \ref{DeltaphiFig} is an estimate from above for the particle diffusion effect. Measuring the value of the corresponding phase shift could allow one to estimate the fraction of the core region where protons are superconducting while neutrons are normal (or moderately superfluid).

\section*{Acknowledgements}
The authors are grateful to Micaela Oertel and J\'er\^ome Novak for discussion. The work was finished during a long-term visit by the authors to the Weizmann Institute of Science (WIS). We acknowledge the support of the visit by the Simons Foundation and WIS. The authors are grateful to the Department of Particle Physics \& Astrophysics at 
WIS for their hospitality and excellent working conditions.
EK and KK also acknowledge Russian Science Foundation [Grant № 19-12-00133] for financial support.

\appendix


\section{Hydrodynamics of superconducting neutron-star matter}\label{AppA}


\subsection{General equations}
\label{general}

In what follows, we use the unit system where the speed of light $c=1$. Although our formulas are valid for the general composition of neutron star cores, for definiteness, we will assume that the matter consists of neutrons $(n)$, protons $(p)$, electrons $(e)$, and muons $(\mu)$, with protons being superconducting, and the rest of the particles being normal (non-superfluid and non-superconducting). Additionally, we will consider the stellar matter to be non-magnetized for simplicity. In this case, the complete system of hydrodynamic equations, accounting for the particle diffusion effects (and neglecting all other dissipative mechanisms), consists of (see Ref. \cite{dg21} for details):

(i) the second law of thermodynamics
\begin{align}
&
{d} \varepsilon = \mu_m {d} n_m + T {d}S + \frac{Y_{pp}}{2} {d}\left(w_{({p})\mu} w^{\mu}_{({p})} \right).
&
\label{2ndlaw}
\end{align}

(ii) energy-momentum conservation
\begin{align}
&
\nabla_\nu T^{\mu\nu}=0,
&
\label{Tmunu0}
\end{align}
where the energy-momentum tensor $T^{\mu\nu}$ equals
\begin{align}
T^{\mu\nu}= (P+\varepsilon) u^{\mu}u^{\nu}+ P g^{\mu\nu}+ Y_{pp} \left(w^\mu_{({p})} 
 w^{\nu}_{({p})}+\mu_p 
 w^{\mu}_{({p})}u^{\nu}+\mu_p w^{\nu}_{({p})} u^{\mu} \right).
&
\label{Tmunu}
\end{align}

(iii) continuity equations for each particle species $m={n}$, ${p}$, ${e}$, and $\mu$
\begin{align}
&
	\nabla_\mu j^\mu_{(m)}=0, 
&
\label{jmu}
\end{align}
where the particle current densities $j^\mu_{(m)}$ are given by
\begin{align}
&	
j^\mu_{(l)}=n_l u^{\mu}+\Delta j^{\mu}_{(l)}, \quad \quad l={n, \, e,\, \mu},
&
\label{jmu1}\\
&
j^\mu_{({p})}=n_{p} u^{\mu} + Y_{pp} w^{\mu}_{({p})} + \Delta j^{\mu}_{(p)},
&
\label{jmu2}
\end{align}
and the diffusion current $\Delta j^{\mu}_{(m)}$ is
\begin{align}
&
\Delta j^{\mu}_{(m)} =- D_{mk} d^{\mu}_{(k)}, \quad\quad 
d^{\mu}_{(k)} \equiv ^\perp\nabla^\mu\left(\frac{\mu_k}{T}\right)-\frac{e_k 
E^{\mu}}{T}.
&
\label{jdiffmy}
\end{align}

(iv) the superfluid equation for protons
\begin{align}
&
\nabla_{\mu}\left(w_{({p})\nu}+e_p A_\nu + \mu_p u_{\nu} \right)-
\nabla_{\nu}\left(w_{({p})\mu}+e_p A_\mu + \mu_p u_{\mu} \right)=0.
&
\label{sfleq2}
\end{align}
In these equations and the following text, space-time indices are represented by Greek letters, while particle species are indicated by indices $m$, $k$, $l$, $q$, and $r$. In all cases, unless explicitly stated otherwise, summation is implied over repeated indices. In equations (\ref{2ndlaw})--(\ref{sfleq2}) $g^{\mu\nu}$ is the metric tensor; $^\perp\nabla_\mu= \perp_{\mu\nu}\nabla^\nu$; $\perp_{\mu\nu}=g_{\mu\nu}+u_{\mu}u_{\nu}$ is the projection tensor; $\nabla_\mu$ is the covariant derivative. Further, $\varepsilon$, $P$, $S$, and $T$ are the energy density, pressure, entropy density and temperature, respectively; $n_m$, $\mu_m$, and $e_m$ are respectively, the number density, the relativistic chemical potential and charge of particle species $m$; $D_{mk}=D_{km}$ are the diffusion coefficients, which can generally be presented as the sole functions of chemical potentials $\mu_m$ and temperature $T$; $u^{\mu}$ is the velocity of normal (nonsuperfluid and nonsuperconducting) liquid component; $A^{\mu}$ is the four-potential of the electromagnetic field; $E^{\mu}\equiv u_\nu F^{\mu\nu}$ is the {\it electric} four-vector, which equals the electric field in the coordinate frame comoving with the normal fluid; $F_{\mu\nu}= \nabla_\mu A_\nu-\nabla_\nu A_\mu$ is the electromagnetic tensor. Next, the coefficient $Y_{pp}$ is the relativistic analogue of the proton superfluid density $\rho_{sp}$ of the nonrelativistic theory. In the nonrelativistic limit they are related by the condition $Y_{pp} = \rho_{\rm sp}/(m_{p}^2 c^2)$, where $c$ is the speed of light and $m_p$ is the bare proton mass. At very low temperatures $T \ll T_{\rm cp}$, where $T_{\rm cp}$ is the proton transition temperature to the superconducting state, all protons condense into Cooper pairs and we have
\begin{align} 
&
\mu_p Y_{pp}= n_{p},
&
\label{Ypp}
\end{align} 
which is analogous to the condition $\rho_{\rm sp}=m_{p} n_{p}$ of the nonrelativistic 
theory.

Finally, the four-vector $w^{\mu}_{({p})}$ 
in equations (\ref{2ndlaw})--(\ref{sfleq2})
is responsible for the superfluid/superconducting degrees of freedom in the system
and is proportional to the difference between the proton superfluid velocity 
$u^{\mu}_{({p})}$ and normal velocity $u^{\mu}$. 
It can be represented as
\begin{align}
&
w^{\mu}_{(p)} = \mu_p (u^{\mu}_{(p)}-u^{\mu}), \quad\quad
u^{\mu}_{({p})} \equiv\frac{1}{\mu_p}\left(\nabla^\mu \phi_{p} -e_{p} A^{\mu} \right),
&
\label{wmu2}
\end{align}
where the scalar $\phi_{p}$ equals $\phi_{p} =\hbar \Phi_{p}/2$ and $\Phi_{p}$ is the wave function phase of the proton Cooper-pair condensate. In fact, equation (\ref{sfleq2}) immediately follows from (\ref{wmu2}) since it just says that $\nabla_\mu \nabla_\nu \phi_{p}-\nabla_\nu \nabla_\mu \phi_{p}=0$.

Multiplying equation (\ref{sfleq2}) by $u^{\nu}$ and using (\ref{wmu2}) one can easily find the expression for the electric vector $E_{\mu}$:
\begin{align}
&
E_\mu=\frac{u^{\nu}}{e_{p}}\left[ \nabla_\nu(\mu_p u_{({p})\mu})-\nabla_\mu(\mu_{p} u_{({p})\nu})  \right].
&
\label{Emu}
\end{align}

The hydrodynamic equations discussed above should be supplemented with a number of additional conditions. The first of these conditions is the normalization of the four-velocity $u^{\mu}$, 
\begin{align}
	&
	u_\mu u^{\mu}=-1,
	&
	\label{wu}
\end{align}
and the condition
\begin{align}
&
u_\mu w^{\mu}_{({p})}=0  \quad \Leftrightarrow \quad u_\mu u^{\mu}_{({p})}=-1,
&
\label{wu}
\end{align}
which ensures that the energy density $\varepsilon$ and the number density $n_{\rm p}$ are measured in the frame comoving with the normal fluid component, i.e., $u_{\mu}u_{\nu} T^{\mu\nu}=\varepsilon$ and $u_\mu j^{\mu}_{(m)}=-n_m$.

The second equation is the expression for the pressure, $P$
\begin{align}
&
P=-\varepsilon + \mu_m n_m +TS,
&
\label{Pres}
\end{align}
and the corresponding Gibbs-Duhem relation
\begin{align}
&
{d} P = n_m {d} \mu_m + S {d}T - \frac{Y_{pp}}{2} {d}\left(w_{(p)\mu} w^{\mu}_{(p)} \right).
&
\label{Giibs}
\end{align}
Finally, the third condition will allow us to close the system of equations described above without invoking Maxwell's equations. This condition follows from the fact that for sufficiently low-frequency motions within the applicability range of the hydrodynamics formulated in this section, the requirement of quasineutrality and vanishing total charge current density in the system is met with a very high degree of accuracy (\cite{kgk21,kgk23}) that is
\begin{align}
&
e_m j^{\mu}_{(m)}=0 \quad \Rightarrow \quad  Y_{pp} w^{\mu}_{({p})}=-\frac{e_m 
\Delta 
j^{\mu}_{(m)}}{e_{p}}.
&
\label{quasineutr}
\end{align}
%


\subsection{Using the fact that $w^{\mu}_{({p})}$ is small}
\label{wpsmall}

As follows from equation (\ref{quasineutr}), the four-vector $w^{\mu}_{({p})}$ depends on small diffusion currents and hence vanishes in full thermodynamic equilibrium. This means that the terms $\sim w^{\mu}_{({p})} w^{\nu}_{({p})}$ in hydrodynamic equations are proportional to the diffusion coefficients squared, $D_{mk}^2$, and, moreover, also depend on the perturbation amplitude squared. In other words, these terms are small and can be neglected either in the case when the diffusion is slow, or if the system is only slightly perturbed out of thermodynamic equilibrium so that use of the linear approximation is justified. Assuming that diffusion is slow equations of Appendix \ref{general} can be reformulated in a very simple and elegant form as
\begin{align}
&
{d} \varepsilon = \mu_m {d} n_m + T {d}S,
\quad\quad
{d} P = n_m {d} \mu_m + S {d}T,
\quad\quad
P=-\varepsilon + \mu_m n_m +TS,
&
\label{2ndlaw2}\\
&
\nabla_\nu T^{\mu\nu}=0, \quad\quad T^{\mu\nu}= (P+\varepsilon) \tilde{u}^{\mu} \tilde{u}^{\nu}+ P 
g^{\mu\nu},
&
\label{Tmunu2}\\
&
\nabla_\mu j^\mu_{(m)}=0, \quad\quad j^\mu_{(m)}=n_m \tilde{u}^\mu +H_{mk} 
\Delta j^{\mu}_{({ k})},
\quad\quad
\Delta j^{\mu}_{(m)} =- D_{mk} d^{\mu}_{(k)}, \quad\quad 
d^{\mu}_{(k)} \equiv \ptnabla^\mu\left(\frac{\mu_k}{T}\right)-\frac{e_k E^{\mu}}{T},
&
\label{jmu2} \\
&
E_\mu=\frac{\tilde{u}^{\nu}}{e_{p}}\left[ \nabla_\nu(\mu_p 
\tilde{u}_{\mu})-\nabla_\mu(\mu_{p} \tilde{u}_{\nu})  \right]
+{\rm small \,\, terms \,\, depending \,\, on \,\, } 
Y_{pp} w^{\mu}_{({p})}.
&
\label{Emu2}
\end{align}
Here we have introduced the differential operator
\begin{gather}
\ptnabla^\mu\equiv \tilde{\perp}^{\mu\nu} \nabla_\nu, \qquad \tilde{\perp}^{\mu\nu} \equiv g^{\mu\nu}+\tilde{u}^\mu\tilde{u}^\nu,
\end{gather}
the matrix $H_{mk}$ is given by ($\delta_{mk}$ is the Kronecker symbol)
\begin{align}
&
H_{mk} =\delta_{mk}-\frac{e_k}{e_p}\left( \delta_{mp}-\frac{\mu_p n_m}{P+\varepsilon}\right),
&
\label{Hmk}
\end{align}
and satisfies 
\begin{align}
&
e_m H_{mk}=0.
&
\label{Hmk1}
\end{align}

The form of the equations has significantly simplified compared to the equations in Appendix \ref{general} because of the introduction of the four-velocity $\tilde{u}^\mu$, which, being multiplied by $(P+\varepsilon)$, gives the momentum density of the fluid:
\begin{align}
&
\tilde{u}^\mu \equiv u^{\mu}+\frac{\mu_p}{P+\varepsilon} \, Y_{pp} w^{\mu}_{({p})}.
&
\label{utilde}
\end{align}
Because of (\ref{wu}), this velocity is normalized to $-1$ with the accuracy to the terms $\sim w_{({p})\mu} w^{\mu}_{({p})}$,
\begin{align}
	&
\tilde{u}_\mu \tilde{u}^\mu=-1.
	&
	\label{utilde2}
\end{align}

Let us note that in writing the expression (\ref{Emu2}) for the electric vector $E_\mu$, we have explicitly written out only the leading terms dependent on the velocity ${\tilde u}^\mu$, as we will neglect contributions from small additional terms depending on $Y_{pp} w^\mu_{({p})}$ (and hence on diffusion coefficients) in what follows (see Appendix \ref{AppB}).

Composing the combination $\tilde{\perp}^{\mu}_\nu \nabla_\lambda T^{\nu\lambda}=0$ and using the expression (\ref{Tmunu2}), one derives the Euler equation 
\begin{align}
	&
	\tilde{u}^\nu \nabla_\nu \tilde{u}^\mu + \tilde{\perp}^{\mu\nu} 
	\frac{\nabla_\nu P}{P+\varepsilon}=0.
	&
\label{eulerx}
\end{align}
Using (\ref{eulerx}), the expression for $E_\mu$ can be transformed into
\begin{align}
	&
E_\mu = \tilde{\perp}^{\nu}_\mu \left(\nabla_\nu \mu_p -\frac{\mu_p}{P+\varepsilon} \nabla_\nu P 
\right).
&
\label{Emu3}
\end{align}
%


\subsection{Hydrodynamic equations for normal matter}

A significant advantage of the hydrodynamic equations in the form obtained 
in Appendix \ref{wpsmall} is their formal resemblance to the corresponding equations for normal 
(non-superfluid and non-superconducting) matter. The complete system of equations for normal matter 
reads as follows:
\begin{align}
	&
	{d} \varepsilon = \mu_m {d} n_m + T {d}S,
	\quad\quad
	{d} P = n_m {d} \mu_m + S {d}T,
	\quad\quad
	P=-\varepsilon + \mu_m n_m +TS,
	&
	\label{2ndlaw2norm}\\
	&
	\nabla_\nu T^{\mu\nu}=0, \quad\quad T^{\mu\nu}= (P+\varepsilon) u^{\mu} u^{\nu}+ P 
	g^{\mu\nu},
	&
	\label{Tmunu2norm}\\
	&
	\nabla_\mu j^\mu_{(m)}=0, \quad\quad j^\mu_{(m)}=n_m u^\mu +H_{mk} \Delta j^{\mu}_{({\rm 
			k})},
	\quad\quad
	\Delta j^{\mu}_{(m)} =- D_{mk} d^{\mu}_{(k)}, \quad\quad 
	d^{\mu}_{(k)} \equiv ^\perp\nabla^\mu\left(\frac{\mu_k}{T}\right)-\frac{e_k E^{\mu}}{T},
	&
	\label{jmu2norm} \\
	&
	E^\mu= - \frac{T e_m D_{mk}}{e_l e_q D_{lq}} \, ^\perp \nabla^\mu \left(\frac{\mu_k}{T}\right).
	&
	\label{Emu2norm}
\end{align}
This system differs from that for matter with superconducting protons only by the expression (\ref{Emu2norm}) for $E^\mu$, which follows from the condition $e_m j^{\mu}_{(m)}=0$. In turn, its superconducting counterpart (\ref{Emu2}) is derived from the superfluid equation for protons, while the requirement $e_m j^{\mu}_{(m)}=0$ is met automatically in view of equation (\ref{Hmk1}). Another (final) difference of the system (\ref{2ndlaw2norm})--(\ref{Emu2norm}) from (\ref{2ndlaw})--(\ref{Hmk}) is the expression for the matrix $H_{mk}$. For normal matter it equals
\begin{align}
&
H_{mk}=\delta_{mk}.
&
\label{Hmknorm}
\end{align}
%



\section{Formulation of the problem in terms of Lagrangian displacements}
\label{AppB}


\subsection{Some definitions}
\label{def}

For a spherically symmetric unperturbed star the interval in $x^{\mu}=(t, r, \theta, \varphi)$ 
coordinates takes the form
\begin{align}
	&
	ds^2=g_{\mu\nu} dx^\mu dx^{\nu}=-{e}^{\nu_0} dt^2 + {e}^{\lambda_0} dr^2+
	r^2 (d\theta^2+{\rm sin}^2\theta d\varphi^2),
	&
	\label{int}
\end{align}
where $\nu_0(r)$ and $\lambda_0(r)$ are (well-known) metric coefficients of the unperturbed star. Here and in the subsequent discussion, unless explicitly specified otherwise, any quantity $A$ pertaining to the unperturbed star is denoted as $A_0$, while the Euler perturbation of this quantity is represented as $\delta A$. Moreover, the superscript $^\infty$ in the quantity $A^\infty$ indicates that this quantity is redshifted as seen by a distant observer, i.e., $A^\infty \equiv A {e}^{\nu_0/2}$.

The tidal force causes a perturbation in the star, resulting in slight variations in the gravitational field
generated by the star. In what follows we shall neglect this variation by working in the so-called Cowling approximation \cite{cowling41}. We will also work in the inertial reference frame, in which the center of mass of the neutron star is at rest each moment of time.
In this reference frame, within made approximations, the effect of the gravitational field of the companion star on the metric tensor can be approximately incorporated by accounting for the corresponding gravitational time dilation: $dt^2 \to (1+2U)dt^2\approx e^{2U} dt^2$, where $U$ is the Newtonian gravitational potential of the companion. Then the metric tensor takes the following form
\begin{align}
	&
	ds^2=-{e}^{\nu_0+2U} dt^2 + {e}^{\lambda_0} dr^2+
	r^2 (d\theta^2+{\rm sin}^2\theta d\varphi^2),
	&
	\label{int2}
\end{align}
which will be used subsequently in writing the Euler equation.
Effectively, such an approach is equivalent to treating the tidal gravitational field
in the Newtonian framework and will lead to the appearance of the tidal force in the Euler equation (see Appendix \ref{operators} below)
\footnote{Such an approach completely ignores the relativistic dragging effects and does not account for the gravitational radiation. These effects, however, are expected to be comparatively small inside the star at large binary orbit separations, considered in the paper.
}
.

In what follows it will be more convenient for us to work with the physical vectors using the physical basis ${\bf e}_\mu$, related to the coordinate basis ${\bf e}_\mu^{\rm coord}\equiv\p_\mu$ as
\begin{align}
	&
	{\pmb{\rm e}}_\mu= S_\mu^{\,\,\,\rho} \pmb{{\rm e}}^{\rm coord}_\rho, \quad \quad 
	S_\mu^{\,\,\rho}={\rm diag}\left[1,{e}^{-\lambda_0/2},\frac{1}{r}, \frac{1}{r \, {\rm 
			sin}\theta}\right].
	&
	\label{phys}
\end{align}
In this new basis the metric is $\eta_{\mu\nu}={\rm diag}[-{e}^{\nu_0+2U},1,1,1]$, while the components of any vector ${\bs A}=A^\mu_{\rm coord}{\bf e}_{\mu}^{\rm coord}=A^\mu {\bf e}_\mu$ in two bases are related by
\begin{align}
	&
	(A^r, A^\theta, A^\varphi)=(A_r, A_\theta, A_\varphi)=({e}^{\lambda_0/2} A^r_{\rm coord},r \, A^\theta_{\rm 
	coord}, 
	r \, {\rm sin}\theta \, A^\varphi_{\rm coord}).
	&
	\label{xi}
\end{align}

Moreover, the gradient, divergence and scalar product in the physical basis are given by 
\begin{align}
	&
	{\pmb \nabla}=\left({e}^{-\lambda_0/2} \frac{\partial }{\partial r}, 
	\frac{1}{r} \frac{\partial }{\partial \theta}, 
	\frac{1}{r {\rm sin}\theta}\frac{\partial }{\partial \varphi}\right),
	&
	\label{grad}\\
	&
	{\rm div} {\pmb A}= \frac{1}{r^2{e}^{\lambda_0/2}} \frac{\partial }{\partial r}
	\left( r^2 A^r\right)
	+ \frac{1}{r {\rm sin}\theta} \frac{\partial }{\partial \theta}
	\left( {\rm sin}\theta A^\theta\right)
	+\frac{1}{r {\rm sin}\theta} \frac{\partial }{\partial \varphi}
	\left( A^\varphi\right), \\
	\label{div}
	&
        ({\pmb A}\cdot{\pmb B})=A^r B^r + A^\theta B^\theta + A^\varphi B^\varphi,
\end{align}
where ${\pmb A}=(A^r,A^\theta,A^{\varphi})$ and  ${\pmb B}=(B^r,B^\theta,B^{\varphi})$ are some vectors with coordinates $A^i$ and $B^i$ in the physical basis ${\bf e}_i$. The volume element is $dV={e}^{\lambda_0/2}r^2 {\rm sin}\theta \, {d}r {d}\theta {d}\varphi$.


\subsection{Our assumptions and equilibrium configuration}

\subsubsection{Assumptions}

Because the stellar matter is strongly degenerate, we will neglect temperature-related contributions in equation (\ref{2ndlaw2norm}) and write
\begin{align}
	&
	{d} \varepsilon = \mu_m {d} n_m,
	\quad\quad
	{d} P = n_m {d} \mu_m,
	\quad\quad
	w \equiv P+\varepsilon = \mu_m n_m,
	&
\label{2ndlaw3norm}
\end{align}
where we define the heat function $w=P+\varepsilon$ (do not confuse with $w_{(p)}^\mu$).

Using these equations, we can present the formula (\ref{Emu3}) for $E_\mu$ as
\begin{align}
&
E_\mu =\left(
\delta_{lp} - \frac{\mu_p n_l}{P+\varepsilon}
\right) \,  ^\perp\tilde{\nabla}_\mu\mu_l.
&
\label{Emu4}
\end{align}

We will also neglect the weak effects in neutron-star matter related to thermal conductivity and 
thermodiffusion. In these circumstances, the kinetic coefficients $D_{mk}$ satisfy the following useful condition \cite{dgs20,dg21},
\begin{align}
	&	
	D_{mk} \mu_{k} =0.
	&
	\label{Dmk}
\end{align}

In the limit of slow diffusion, considered here, diffusive currents $\Delta j^\mu_{(m)}$ are given by Eq.\ \eqref{jmu2}. Using equations (\ref{Emu4}) and (\ref{Dmk}), they can be transformed into the following form
\begin{align}
&
\Delta j^{\mu}_{(m)} =- \frac{D_{mk} H_{lk}}{T} \, ^\perp \tilde{\nabla}^\mu \mu_l,
&
\label{jmu4}
\end{align}
where the matrix $H_{lk}$ is given by the formula (\ref{Hmk}). Note that,  because of (\ref{2ndlaw3norm}) this matrix has a property
\begin{align}
&
\mu_m H_{mk}=\mu_k
&
\label{Hmk2}
\end{align}
in addition to the property (\ref{Hmk1}).
In view of (\ref{Dmk}) and (\ref{Hmk2}), the expression (\ref{jmu4})
can be rewritten as
\begin{align}
	&
	\Delta j^{\mu}_{(m)} =- \frac{D_{mk} H_{lk}}{T^\infty} \, ^\perp \tilde{\nabla}^\mu 
	\mu_l^\infty,
	&
	\label{jmu44}
\end{align}
where $T^\infty=T {e}^{\nu_0/2}$ and $\mu_l^\infty=\mu_l {e}^{\nu_0/2}$.

\subsubsection{Equilibrium configuration of a neutron star}

In an unperturbed star, one has (see, e.g., Refs.\ \cite{ga06,gkcg13})
\begin{align}
&
u^{\mu}_0 = ({e}^{-\nu_0/2},0,0,0),
\quad\quad
w^{\mu}_{({p})0}=(0, 0,0, 0),
\quad \quad
\tilde{u}^{\mu}_0 = ({e}^{-\nu_0/2},0,0,0).
&
\label{u0}
\end{align}
Moreover, in the unperturbed star the redshifted internal temperature must be constant
and the diffusion current must vanish, which imply (see equations \ref{jmu2} and \ref{u0})
\begin{align}
&
{\pmb \nabla} T_0^\infty =0,
\quad\quad
{\pmb \nabla}\mu_{k0}^\infty -e_k {\pmb E}_0 \, e^{\nu_0/2} =0.
&
\label{equil}
\end{align}
Here the gradient $\pmb{\nabla}$ is defined in equation (\ref{grad}) and ${\pmb E}_0$ is the spatial part of the electric four-vector $E^{\mu}=u_\nu F^{\mu\nu}$ in the unperturbed star. From (\ref{equil}) it is straightforward to show that
\begin{align}
	&
	e_m {\pmb \nabla}\mu_{k0}^\infty -e_k {\pmb \nabla}\mu_{m0}^\infty =0
	&
	\label{muT2}
\end{align}
for any two arbitrary species $m$ and $k$.


\subsection{Operators $\hat{\mathcal{L}}$ and $\hat{\mathcal{D}}$}
\label{operators}

The hydrodynamic velocity $\tilde{{\bs u}}$ in an arbitrary reference frame can be written as $\tilde{{\bs u}}=\tilde{{\bs u}}_{\rm NS}+(\tilde{{\bs u}}-\tilde{{\bs u}}_{\rm NS})=\tilde{{\bs u}}_{\rm NS}+\delta\tilde{\bs u}$, where $\tilde{{\bs u}}_{\rm NS}$ is the neutron star center of mass velocity (coinciding with the equilibrium hydrodynamic velocity). Here we employ the inertial reference frame, comoving with the center of mass of the neutron star. In this frame $\tilde{{\bs u}}_{\rm NS}=0$ and the metric tensor takes the form \eqref{int2}. One can also show that in this frame the center of mass acceleration equals $\dot{\tilde{\bs u}}_{\rm NS}=-e^{\nu_0/2}({\bs\nabla} U)_{l=1}=e^{\nu_0/2}GM'{\bs D}/D^3$. In this frame, linearizing equation (\ref{eulerx}) and introducing the Lagrangian displacement $\bs\xi$ as $\delta\tilde{{\bs u}}=e^{-\nu_0/2}\dot{\bs\xi}$, we obtain the following Euler equation for stellar perturbations under the action of tidal potential (written in the physical basis):
\begin{align}
	&
	{e}^{-\nu_0} w_0 \, \ddot{{\bs \xi}} = -{\pmb \nabla}\delta P + \frac{\delta w}{w_0}{\pmb 
		\nabla} 
	P_0 -w_0 {\pmb \nabla}\tilde{U}.
	&
	\label{Eu}
\end{align}
In turn, the linearized version of the continuity equation (\ref{jmu2}) will take the form
\begin{align}
	&
	\dot{\delta n_m}+ {\rm div} (n_{m0} \dot{{\bs \xi}})=-  {\rm 
		div} \left( H_{mk0} \Delta {\pmb j}_{k} \, {e}^{\nu_0/2}\right).
	&
	\label{cont}
\end{align}
In these equations dot ($\dot{\,\,}$) indicates time derivative $\partial/\partial t$; $\Delta {\pmb j}_{k}$ is the (linearized) spatial components of the diffusion current (\ref{jmu44}) (note that it vanishes in equilibrium); ${\pmb \nabla}$ and ${\rm div}$ are given, respectively, by equations (\ref{grad}) and (\ref{div}).

Because of the conditions (\ref{Hmk1}), (\ref{Dmk}), (\ref{Hmk2}), and (\ref{muT2}) $\Delta {\pmb j}_{k}$ can be written as follows
\begin{align}
	&
	\Delta {\pmb j}_{m} =
- \frac{D_{mk0} H_{lk0}}{T^\infty} 
{\pmb \nabla }
	\delta\mu_l^\infty,
	&
	\label{jmu5}
\end{align}
where the matrices $D_{mk0} = D_{mk}(\{\mu_{q0}\}, T)$ and $H_{lk0} = H_{lk}(\{\mu_{q0}\})$ are taken at equilibrium (more precisely, at the equilibrium values of chemical potentials $\mu_{q0}$; $q=n$, $p$, $e$, $\mu$). However, it should be noted that to derive this formula, the true temperature $T$ does not necessarily need to be equal (or close) to the equilibrium temperature $T_0$. In other words, Eq.\ (\ref{jmu5}) remains valid even for significant deviations from thermal equilibrium, when $T$ differs noticeably from $T_0$ (but still $\delta \mu_q \ll \mu_{q0}$). In the case of strong deviations from thermal equilibrium, it is necessary to substitute the true temperature $T$ into the temperature-dependent coefficients $D_{mk0}$.

Using one of the hydrostatic equilibrium conditions (e.g., \cite{hpy07}), ${\pmb \nabla}P_0 = - w_0 {\pmb \nabla} \nu_0/2$, equation (\ref{Eu}) can be rewritten as
\begin{align}
	&
	{e}^{-\nu_0/2} w_0 \, \ddot{{\bs \xi}} = -{\pmb \nabla}\left(\delta P {e}^{\nu_0/2}\right) -\delta \varepsilon {\pmb \nabla}\left({e}^{\nu_0/2}\right)
	-w_0\, {e}^{\nu_0/2}\, {\pmb \nabla}\tilde{U}(t, {\pmb r}).
	&
	\label{Eu2}
\end{align}

Our next aim will be to express Euler perturbations of pressure, $\delta P$, chemical potentials, $\delta \mu_k$, and energy density, $\delta \varepsilon$, from the continuity equations (\ref{cont}). Using (\ref{2ndlaw3norm}) as well as the identity $\delta P= (\partial P/\partial n_m) \, \delta n_m$ one arrives at the following set of equations (see also \cite{kgk23})
\begin{align}
	&
	\dot{\delta P}+ {\pmb \nabla}P_0 \cdot \dot{\bs{\xi}} + n_{m0} \frac{\partial P}{\partial n_m} 
	\, {\rm div} \dot{{\bs 
			\xi}}=-  
	\frac{\partial P}{\partial n_m}{\rm 
		div} \left(H_{mk0} \Delta {\pmb j}_{k} \, {e}^{\nu_0/2}\right),
	&
	\label{dP}
\end{align}
\begin{align}
	&
	\dot{\delta \mu}_m+ {\pmb \nabla}\mu_{m0} \cdot \dot{\bs{\xi}} + \frac{\partial P}{\partial 
		n_m} {\rm div} \dot{{\bs 
			\xi}}=-  
	\frac{\partial \mu_m}{\partial n_l}{\rm 
		div} \left(H_{lk0} \Delta {\pmb j}_{k} \, {e}^{\nu_0/2}\right),
	&
	\label{dmuk}
\end{align}
\begin{align}
	&
	\dot{\delta \varepsilon}+ {\pmb \nabla}\varepsilon_0 \cdot \dot{\bs{\xi}} + w_0 \, {\rm div} 
	\dot{{\bs 
			\xi}}=-  
	\mu_{m0}{\rm 
		div} \left(H_{mk0}\Delta {\pmb j}_{k} \, {e}^{\nu_0/2}\right).
	&
	\label{eps}
\end{align}
where all partial derivatives of thermodynamic quantities are taken in the unperturbed star;
it is assumed that the pressure and chemical potentials are presented as the functions of number 
densities of all particle species.

In view of the identities (\ref{Hmk1}), (\ref{Dmk}), (\ref{Hmk2}), (\ref{muT2}), and (\ref{jmu5})
the right-hand side of equation (\ref{eps}) vanishes, hence we are left with
\begin{align}
	&
	\dot{\delta \varepsilon}+ {\pmb \nabla}\varepsilon_0 \cdot \dot{\bs{\xi}} + w_0 \, {\rm div} 
	\dot{{\bs 
			\xi}}=0.
	&
	\label{eps2}
\end{align}

Now, let us formally present $\delta P$ as $\delta P=\delta P_{\rm ad}+\delta P_{\rm diff}$, where $\delta P_{\rm ad}$ is expressed through $\bs\xi$ by the same formula as in the absence of diffusion (see equation \ref{dP} with the zero right-hand side) and $\delta P_{\rm diff}$ is the diffusion correction arising because the right-hand side of \ref{dP} is non-vanishing
\footnote{Similar representation for the energy density perturbation gives $\delta \varepsilon= \delta \varepsilon_{\rm ad}$ because $\delta \varepsilon_{\rm diff}=0$, see equation (\ref{eps2}). }.
Then, substituting $\delta P=\delta P_{\rm ad}+\delta P_{\rm diff}$ and $\delta \varepsilon= \delta \varepsilon_{\rm ad}$ into (\ref{Eu2}), we get
\begin{align}
	&
	{e}^{-\nu_0/2} w_0 \, \ddot{{\bs \xi}} =\hat{\mathcal{L}}{\bs \xi} -{\pmb 
	\nabla}\left(\delta P_{\rm 
	diff} {e}^{\nu_0/2}\right) 
	-w_0\, {e}^{\nu_0/2}\, {\pmb \nabla}\tilde{U}(t, {\pmb r}).
	&
	\label{Eu22}
\end{align}
where the operator $\hat{\mathcal{L}}$ is defined as
\begin{align}
	&
	\hat{\mathcal{L}}{\bs \xi} = {\pmb \nabla}\left[
	\left(
	{\pmb \nabla}P_0 \cdot {\bs \xi} + n_{m0}\frac{\partial P}{\partial n_m} \, {\rm div}{\bs \xi}
	\right) {e}^{\nu_0/2}
	\right] 
	+\left(
	{\pmb \nabla} \varepsilon_0 \cdot {\bs \xi} + w_0 \, {\rm div} {\bs \xi}
	\right) {\pmb \nabla}\left({e}^{\nu_0/2} \right).
	&
	\label{Lxi}
\end{align}
One can show that for any functions ${\bs\xi}_1$ and ${\bs\xi}_2$, satisfying the standard boundary conditions at the surface (see Sec.\ \ref{perturb}), this operator is symmetric in a sense that
\begin{gather}
\label{symmetric}
\int {\bs\xi}_1 \hat{\mathcal{L}} {\bs\xi}_2 dV=\int {\bs\xi}_2 \hat{\mathcal{L}} {\bs\xi}_1 dV, \qquad dV=e^{\lambda_0/2}r^2\sin\theta dr d\theta d\varphi, 
\end{gather}
where the integration is performed over the whole stellar volume.

In turn, as follows from equation (\ref{dP}), the {\it time-derivative} of the diffusion correction to pressure $\delta P_{\rm diff}$ is given by
\begin{align}
&
\dot{\delta P}_{\rm diff}=-  
\frac{\partial P}{\partial n_m}{\rm 
	div} \left(H_{mk0} \Delta {\pmb j}_{k} \, {e}^{\nu_0/2}\right).
&
\label{dPdiffu}
\end{align}
The function $\delta P_{\rm diff}$ itself can be found from (\ref{dPdiffu}) by integration over time. Using equation (\ref{jmu5}), the formula (\ref{dPdiffu}) can be represented as
\begin{align}
	&
	\dot{\delta P}_{\rm diff}=
	\frac{\partial P}{\partial n_m}{\rm 
		div} \left(
\, {e}^{\nu_0/2} 
\frac{H_{mk0}  D_{kq0} H_{lq0}}{T^\infty} 
{\pmb \nabla }
\delta\mu_l^\infty
\right)
&
\nonumber\\
&
\approx 
-\frac{\partial P}{\partial n_m}{\rm 
	div} \left[
\, {e}^{\nu_0/2}
\frac{H_{mk0}  D_{kq0} H_{lq0}}{T^\infty} 
{\pmb \nabla }
\left( {e}^{\nu_0/2}
\frac{
\partial P}{\partial n_l}  {\rm div} {\bs \xi}
\right)
\right].
&
	\label{dPdiffu2}
\end{align}
Here in the last equality we expressed $\delta \mu_l$ from equation (\ref{dmuk}). In doing so, we retained only the terms linear in the diffusion coefficients $D_{kq0}$ (assuming diffusion is slow) and utilized the properties (\ref{Hmk1}) and (\ref{Hmk2}).

Now we have everything at hand to derive the energy conservation law from (\ref{Eu22}). Multiplying that equation by $\dot{{\bs \xi}}$, integrating over the stellar volume $V$, and using the property (\ref{symmetric}) one finds, after some manipulations,
\begin{align}
	&
\frac{d}{dt} \int {d}V \left[ 
\frac{1}{2} \, {e}^{-\nu_0/2} \, w_0 \, \dot{\bs \xi}^2 - \frac{1}{2} {\bs \xi} 
\hat{\mathcal{L}}{\bs \xi} + 
{\bs \xi} {\pmb 
	\nabla}\left(\delta P_{\rm diff} {e}^{\nu_0/2}\right) \right]
=  \int {d} V {\bs \xi} {\pmb 
	\nabla}\left(\dot{\delta P}_{\rm diff} {e}^{\nu_0/2}\right)   -\int {d}V w_0\, {\rm 
	e}^{\nu_0/2}\, 
	\dot{\bs \xi}\,{\pmb \nabla}\tilde{U}(t, {\pmb r}).
&
\label{energy0}
\end{align}
or, making use of (\ref{dPdiffu2}),
\begin{align}
	&
\frac{d}{dt} \int {d}V \left[ 
\frac{1}{2} \, {e}^{-\nu_0/2} \, w_0 \, \dot{\bs \xi}^2 - \frac{1}{2} {\bs \xi} 
\hat{\mathcal{L}}{\bs \xi} + 
{\bs \xi} {\pmb 
	\nabla}\left(\delta P_{\rm diff} {e}^{\nu_0/2}\right) \right]
= \int {d}V {\bs \xi} \hat{\mathcal{D}} {\bs \xi} - \int {d}V w_0\, {e}^{\nu_0/2}\, 
	\dot{\bs \xi}\, {\pmb \nabla}\tilde{U}(t, {\pmb r}),
&
\label{energy}
\end{align}
where
\begin{align}
	&
	\hat{\mathcal{D}}{\bs \xi} \equiv -
	{\pmb \nabla}\left\{
	{e}^{\nu_0/2} \frac{\partial P}{\partial n_m} {\rm div}\left[
	\frac{	H_{mk0}  D_{kq0} H_{lq0}}{T} 
	{\pmb \nabla }
	\left( 
	{e}^{\nu_0/2} \frac{\partial P}{\partial n_l} {\rm div}{\bs \xi}
	\right)
	\right]
	\right\}.
	&
	\label{Dxi2}
\end{align}
One can show that the operator $\hat{\mathcal{D}}$ has the following properties:
\begin{gather}
\label{Dsym}
\qquad \int {\bs\xi}_1\hat{\mathcal{D}}{\bs\xi}_2 dV=\int {\bs\xi}_2\hat{\mathcal{D}}{\bs\xi}_1 dV, \qquad  \int {\bs\xi}\hat{\mathcal{D}}{\bs\xi} dV\leq 0.
\end{gather}
The first property is obtained by integrating $\int {d}V {\bs \xi}_1\hat{\mathcal{D}}{\bs \xi}_2$ a few times by parts. The only assumption required for it to hold is the vanishing of the diffusion coefficients $D_{km}$ and their derivatives at the surface. The second property, as can be shown, follows from the non-negative-definiteness of the matrix $D_{km}$ \cite{dgs20,dg21}.

The equation (\ref{energy}) has a straightforward physical meaning and indicates that the change in the total mechanical energy of the system over time is due to the dissipation of energy into heat through diffusion (the first term on the right-hand side of equation \ref{energy}) and the work done on the system by the tidal force (the second term on the right-hand side of the equation). In turn, the mechanical energy consists of kinetic energy (the first term on the left-hand side of the equation), potential energy of a standard form (the second term), and a small addition to the internal energy caused by the difference in pressure between the matter where diffusion is active and the matter where diffusion is artificially turned off and the composition is ``frozen'' (the third term on the left-hand side of the equation \ref{energy}). Notably, when oscillations are harmonic (neglecting dissipation) time-averaging of the third term vanishes up to the leading terms in diffusion. 

To end this section we would like to note that $\dot{E}_{\rm diff}$ can be equivalently written in terms of chemical potential perturbations. To show this, as a first step, we twice integrate ${\bs\xi}\hat{\mathcal{D}}{\bs\xi}$ by parts and obtain:
\begin{gather}
\dot{E}_{\rm diff}=\int {\bs\xi}\hat{\mathcal{D}}{\bs\xi} dV = -\int \frac{1}{T}H_{mk0}H_{lq0}D_{kq0}{\pmb\nabla}\biggl[e^{\nu_0/2}\frac{\p P}{\p n_m}\divv{\bs \xi}\biggr]\cdot{\pmb\nabla}\biggl[e^{\nu_0/2}\frac{\p P}{\p n_l}\divv{\bs \xi}\biggr] dV.
\end{gather}
Then we substitute $\divv{\bs \xi}$ from the nondissipative limit of the equation \eqref{dmuk} and get
\begin{gather}
\dot{E}_{\rm diff}=-\int \frac{1}{T}H_{mk0}H_{lq0}D_{kq0}{\pmb\nabla}[\delta\mu_m^\infty+e^{\nu_0/2}({\pmb \xi{\pmb\nabla}})\mu_{m0}]\cdot{\pmb\nabla}[\delta\mu_l^\infty+e^{\nu_0/2}({\pmb \xi{\pmb\nabla}})\mu_{l0}] dV.
\end{gather}
It is easy to see that terms with ${\pmb\nabla}\mu_{m0}$ and ${\pmb\nabla}\mu_{l0}$ vanish. Indeed, using that
\begin{gather}
H_{mk0} e_m=0, \quad \mu_{m0}H_{mk0}=\mu_{k0}, \quad \mu_{k0}D_{kq0}=0, \quad \mu_{m0}^\infty-e_m/e_{p}\mu_{p 0}^\infty={\rm const},
\end{gather}
we find that
\begin{gather}
H_{mk0}D_{kq0}{\pmb\nabla}[e^{\nu_0/2}({\bs \xi \pmb\nabla})\mu_{m0}]=0.
\end{gather}
Thus, $\dot{E}_{\rm diff}$ can be written as
\begin{gather}
\dot{E}_{\rm diff}=-\int \frac{1}{T}H_{mk0}H_{lq0}D_{kq0}({\pmb\nabla}\delta\mu_m^\infty)\cdot({\pmb\nabla}\delta\mu_l^\infty) dV.
\end{gather}
%



\section{Evolution of the binary system in Newtonian theory}\label{AppC}

Let us consider a binary system, consisting of a neutron star with mass $M_{\rm NS}$ and its point-mass companion with mass $M_{\rm C}$. Let the vector ${\bs R}_{\rm C}(t)$ describe the trajectory of the companion, and the vector ${\bs R}_{\rm NS}(t)$ describe the trajectory of the neutron star's center of mass:
\begin{gather}
\label{cmNS}
{\bs R}_{\rm NS}(t)=\frac{1}{M_{\rm NS}}\int \rho({\bs R}-{\bs R}_{\rm NS}(t),t){\bs R} \, d^3R, \qquad M_{\rm NS}=\int \rho({\bs R}-{\bs R}_{\rm NS}(t),t) \, d^3R,
\end{gather}
where $\rho({\bs R}-{\bs R}_{\rm NS},t)$ is the density of the neutron star as a function of the distance ${\bs r}={\bs R}-{\bs R}_{\rm NS}$ from its center of mass. In what follows, we use $\bs R$ to denote the positions of points with respect to an arbitrarily chosen inertial reference frame. Velocities of the companion and the neutron star center of mass, measured in this reference frame, equal $\dot{\bs R}_{\rm C}$ and $\dot{\bs R}_{\rm NS}$, respectively. Let us also introduce the velocity ${\bs V}$ of hydrodynamical flows inside the star, measured in the same inertial frame. By definition, it is related to the ordinary hydrodynamical velocity $\bs v$ as ${\bs V}=\dot{\bs R}_{\rm NS}+{\bs v}$. It is easy to show that, in these notations, the kinetic energy of the binary system can be written as
\begin{gather}
E_{\rm kin}=\frac{1}{2}M_{\rm C}\dot{{\bs R}}_{\rm C}^2+\frac{1}{2}\int \rho({\bs R}-{\bs R}_{\rm NS},t){\bs V}^2 \, d^3R=\frac{1}{2}M_{\rm C}\dot{{\bs R}}_{\rm C}^2+\frac{1}{2}M_{\rm NS}\dot{{\bs R}}_{\rm NS}^2+\frac{1}{2}\int\rho{\bs v}^2 \,d^3r.
\end{gather}
In the derivation of this formula we have used that
\begin{gather}
\label{intRhoR}
\int \rho({\bs r},t){\bs r} \, d^3r=0, \qquad \int \rho({\bs r},t) {\bs v}({\bs r},t) \, d^3r=0,
\end{gather}
which follows from \eqref{cmNS} and the continuity equation $\dot{\rho}+\divv(\rho{\bs v})=0$. The obtained formula for kinetic energy is intuitively clear: the first two terms correspond to the motion of the companion and the neutron star as a whole, while the last term describes the kinetic energy of hydrodynamical flows inside the star. Potential energy, in turn, represents the sum of the interaction energy between the neutron star and its companion and hydrodynamical potential energy $E_{\rm pot, hydro}$:
\begin{gather}
E_{\rm pot}=-GM_{\rm C}\int \frac{\rho({\bs r},t) \, d^3 r}{|{\bs r}-({\bs R}_{\rm C}-{\bs R}_{\rm NS})|}+E_{\rm pot, hydro}.
\end{gather}
Since we consider weak perturbations of a nonrotating neutron star, hydrodynamic contributions to kinetic and potential energy can be written as
\begin{gather}
E_{\rm kin, hydro}=\frac{1}{2}\int\rho_0(r) \dot{{\bs \xi}}^2 \,d^3r, \qquad E_{\rm pot, hydro}=-\frac{1}{2}\int{\bs\xi}\hat{\mathcal{L}}{\bs\xi} \, d^3r,
\end{gather}
where $\rho_0$ is the unperturbed density. As a result, in this limit we have
\begin{gather}
E_{\rm kin}=\frac{1}{2}M_{\rm C}\dot{{\bs R}}_{\rm C}^2+\frac{1}{2}M_{\rm NS}\dot{{\bs R}}_{\rm NS}^2+\frac{1}{2}\int\rho_0(r) \dot{{\bs \xi}}^2 \,d^3r, \\
E_{\rm pot}=-GM_{\rm C}\int \frac{\rho({\bs r},t)}{|{\bs r}-({\bs R}_{\rm C}-{\bs R}_{\rm NS})|}\, d^3 r-\frac{1}{2}\int{\bs\xi}\hat{\mathcal{L}}{\bs\xi} \, d^3r.
\end{gather}
Further analysis is more transparent, if we separate the motion into the motion of the binary as a whole and the relative motion of the neutron star and its companion. To this aim, let us introduce the center of mass ${\bs R}_{\rm CM}$ of the whole binary system and distance ${\bs D}$ between the neutron star and companion:
\begin{gather}
{\bs R}_{\rm CM}=\frac{M_{\rm NS}{\bs R}_{\rm NS}+M_{\rm C}{\bs R}_{\rm C}}{M_{\rm NS}+M_{\rm C}}, \qquad {\bs D}={\bs R}_{\rm C}-{\bs R}_{NS}.
\end{gather}
Using these relations one can rewrite kinetic and potential energies in terms of ${\bs R}_{\rm CM}$ and ${\bs D}$ as
\begin{gather}
E_{\rm kin}=\frac{1}{2}(M_{\rm C}+M_{\rm NS})\dot{{\bs R}}_{\rm CM}^2+\frac{1}{2}\mu \dot{{\bs D}}^2+\frac{1}{2}\int\rho_0(r) \dot{{\bs \xi}}^2 \,d^3r, \qquad \mu=\frac{M_{\rm C}M_{\rm NS}}{M_{\rm C}+M_{\rm NS}}, \\
E_{\rm pot}=\int\rho({\bs r},t)U({\bs r},{\bs D})\, d^3 r-\frac{1}{2}\int{\bs\xi}\hat{\mathcal{L}}{\bs\xi} \, d^3r, \qquad U({\bs r},{\bs D})=-\frac{GM_{\rm C}}{|{\bs r}-{\bs D}|}.
\end{gather}
The total energy of the binary, therefore, can be written as
\begin{gather}
E_{\rm tot}=E_{\rm kin}+E_{\rm pot}=E_{\rm kin, CM}+E_{\rm orb}+E_{\rm mech}, \\
E_{\rm kin, CM}\equiv\frac{1}{2}(M_{\rm C}+M_{\rm NS})\dot{{\bs R}}_{\rm CM}^2, \\
E_{\rm orb}\equiv\frac{1}{2}\mu \dot{{\bs D}}^2+\int\rho({\bs r},t)U({\bs r},{\bs D})\, d^3 r, \\
E_{\rm mech}\equiv\frac{1}{2}\int\biggl[\rho_0(r) \dot{{\bs \xi}}^2-{\bs\xi}\hat{\mathcal{L}}{\bs\xi}\biggr] \, d^3r.
\end{gather}
Here $E_{\rm kin, CM}$ is the kinetic energy, associated with the motion of the binary's center of mass, $E_{\rm orb}$ is the orbital energy, containing the gravitational interaction energy and kinetic energy of relative motion of the star and companion, and $E_{\rm mech}$ is the mechanical energy of neutron star perturbations.

Once the expressions for the kinetic and potential energies are obtained, one can write the action $\mathcal{S}$ for the binary system. In doing so one should remember that the variables ${\bs R}_{\rm NS}$ (and, therefore $\bs D$) and $\bs\xi$ are not independent and are bound by condition \eqref{cmNS}, which is equivalent to the first equation in \eqref{intRhoR}. These equations are nothing but the definition of the radial coordinate $\bs r$, claiming that the center of mass of the neutron star is always located at ${\bs r}=0$ (in particular, in the perturbed star). In other words, stellar perturbations should leave the center of mass at rest with respect to the chosen reference frame. This condition imposes the following restriction on the perturbations,
\begin{gather}
\label{cmpos0}
\int \delta\rho {\bs r}d^3r=0,
\end{gather}
which is the first linearized equation in \eqref{intRhoR}. Then the action of the binary system, accounting for this restriction, can be written as
\begin{gather}
\mathcal{S}_{\bs\kappa}\{{\bs R}_{\rm CM}, {\bs D},{\bs\xi}\}=\int(E_{\rm kin}-E_{\rm pot})dt+{\bs\kappa}\int\delta\rho \, {\bs r} \, d^3r \, dt,
\end{gather}
where $\bs\kappa$ is a Lagrangian multiplier, introduced to enforce \eqref{cmpos0}. Equations, governing the dynamics of the binary system, can be derived by minimizing this action with respect to variations of ${\bs R}_{\rm CM}$, ${\bs D}$, $\bs\xi$. The Lagrangian multiplier can be found by requiring the condition \eqref{cmpos0}. We would like to note that, in our approach, density perturbation is not independent and equals $\delta\rho=-\divv(\rho_0{\bs\xi})$ (which is the linearized continuity equation). This relation should be accounted for in the minimization of the action with respect to variations of ${\bs\xi}$. Keeping this in mind, we find that minimization with respect to variations of $\bs\xi$ and condition \eqref{cmpos0} lead to the following equations
\begin{gather}
\rho_0\ddot{\bs\xi}-\hat{\mathcal{L}}{\bs\xi}={\bs \kappa}\rho_0-\rho_0{\pmb\nabla} U, \label{first} \\
\int \rho_0{\bs\xi}d^3r=0. \label{second}
\end{gather}
The Lagrangian multiplier $\bs\kappa$ can be found by integrating Eq.\ (\ref{first}) and using Eq.\ (\ref{second}) (note that $\int \hat{\mathcal{L}}\bs \xi dV$ vanishes due to Eq.\ \ref{second} and the fact that the operator $\hat{\mathcal{L}}$ conserves the angular dependence of $\bs \xi$). Excluding ${\bs\kappa}$ from the resulting equation and inserting it into Eq.\ (\ref{first}), we find 
\begin{gather}
\rho_0\ddot{\bs\xi}-\hat{\mathcal{L}}{\bs\xi}=-\rho_0[\pmb\nabla U-\pmb\nabla U_{l=1}]=-\rho_0 \pmb\nabla \tilde{U}.
\end{gather}
Detailed analysis of this equation (more specifically, of its relativistic dissipative counterpart) has already been performed in the main text of the paper. Similar consideration allows one to derive the conservation law for mechanical energy:
\begin{gather}
\dot{E}_{\rm mech}=-\int\limits \rho_0(\dot{\bs\xi} \pmb\nabla)\tilde{U}d^3r\equiv A_U,
\end{gather}
where $A_U$ is the work of gravitational forces. Further, minimization of the action with respect to variations of ${\bs D}$ gives
\begin{gather}
\mu \ddot{\bs D}=-\int \rho({\bs r},t) \frac{\p U}{\p {\bs D}} d^3r.
\end{gather}
Multiplying this equation by $\dot{\bs D}$ and rearranging the terms, we find
\begin{gather}
\dot{E}_{\rm orb}=\int U\frac{\p\rho}{\p t} \, d^3r=\int U\frac{\p\delta\rho}{\p t} \, d^3r,
\label{Eorbdot}
\end{gather}
Substituting explicitly $\delta\rho$ expressed through $\bs\xi$ and integrating the right-hand side by parts, we obtain
\begin{gather}
\dot{E}_{\rm orb}=\int\limits \rho_0(\dot{\bs\xi}\pmb\nabla) U d^3r=\int\limits \rho_0(\dot{\bs\xi}\pmb\nabla)\tilde{U} d^3r=-A_U,
\end{gather}
where we have used the time derivative of Eq.\ (\ref{second}) and that $\pmb\nabla U_{l=1}$ does not depend on $\bs r$. Finally, minimization of the action with respect to ${\bs R}_{\rm CM}$ gives $\ddot{\bs R}_{\rm CM}=0$. This is the expected result, since in the absence of external forces, the center of mass of the isolated system should move with constant velocity.


\section{Effect of diffusion on the tidal perturbation}\label{AppD}

In the text we have mentioned that diffusion results in delay of the perturbation in time and, moreover, strong diffusion suppresses perturbation amplitude. 
To show this, let us consider the Euler equation, accounting for diffusion this time. It is easy to show that, operating in the same manner as in the main text, one arrives at the following equation for $b_{\bf k}$:
\begin{gather}
\ddot{b}_{\bf k}(t)-2im\Omega\dot{b}_{\bf k}(t)+(\omega_{\bf k}^2-m^2\Omega^2-im\dot{\Omega})b_{\bf k}(t)+\sum_{\bf k'}\int\limits_{-\infty}^t \gamma_{\bf k k'}(\tau) b_{\bf k'}(\tau)
e^{-im'\Phi(\tau)+im\Phi(t)}d\tau=\frac{1}{D(t)^{l+1}}, \\
\gamma_{\bf k k'}\equiv\frac{Q_{\bf k'} W_{\bf k'}}{Q_{\bf k} W_{\bf k}}\int {\pmb\xi}_{\bf k}^\star \hat{\mathcal{D}} {\bs\xi}_{\bf k'} dV.
\end{gather}
As in the case of $\dot{E}_{\rm diff}$ \eqref{Ediffak}, the only nonvanishing cross-terms in the sum above are those with $l'=l$ and $m'=m$. 
We will focus on the $m\neq 0$ case, because $m=0$ modes do not contribute either to GW emission or to diffusive dissipation. Then we can approximate integrals in the sum as
\begin{gather}
\int\limits_{-\infty}^t \gamma_{\bf k k'}(\tau) b_{\bf k'}(\tau)
e^{-im(\Phi(\tau)-\Phi(t))}
d\tau\approx\gamma_{\bf k k'}(t) b_{\bf k}(t) \lim_{\delta\to +0}\int\limits_{-\infty}^t 
e^{-im(\Phi(\tau)-\Phi(t))(1+i\delta)}
d\tau=\frac{i}{m \dot{\Phi}(t)}\gamma_{\bf k k'}(t)b_{\bf k'}(t).
\end{gather}
Here in order to take the integral from the exponent we introduced a small correction $\delta\to +0$ to regularize the integral at $\tau\to -\infty$. Physically such regularization reflects the idea that at $\tau\to-\infty$ the neutron star is not perturbed by its companion. Then the equation for $b_{\bf k}$ can be rewritten as
\begin{gather}
\ddot{b}_{\bf k}-2im\Omega\dot{b}_{\bf k}+(\omega_{\bf k}^2-m^2\Omega^2-im\dot{\Omega})b_{\bf k}+\frac{i}{m\Omega}\sum_{\bf k'}\gamma_{\bf k k'} b_{\bf k'}=\frac{1}{D^{l+1}}.
\end{gather}
At large orbit separations $b_{\bf k}$ can be found from the asymptotic system of equations:
\begin{gather}
\sum_{\bf k'}\biggl[\delta_{\bf k k'}+\frac{i}{m\Omega}\frac{\gamma_{\bf k k'}}{\omega_{\bf k}^2-m^2\Omega^2}\biggr]b_{\bf k'}=\frac{1}{D^{l+1}(\omega_{\bf k}^2-m^2\Omega^2)}.
\end{gather}
To qualitatively analyze the behaviour of the solution of this system of equations, let us discard the cross-terms and assume that the mode is far from the resonance. Then we obtain
\begin{gather}
\biggl[1-\frac{i}{m\Omega \tau_{\rm diff,\,\bf k}}\biggr]b_{\bf k}=\frac{1}{D^{l+1}\omega_{\bf k}^2},
\end{gather}
where
$\tau_{\rm diff,\,\bf k}$ is defined by Eq.\ (\ref{tauDiff}). The solution reads as:
\begin{gather}
b_{\bf k}=\frac{1}{D^{l+1}\omega_{\bf k}^2}\frac{1}{1+\frac{1}{m^2\Omega^2 \tau^2_{\rm diff,\,\bf k}}}\biggl[1+\frac{i}{m\Omega \tau_{\rm diff,\,\bf k}}\biggr],
\end{gather}
which implies that diffusion introduces the phase delay to the perturbation%
\footnote{
For weak diffusion, this delay is small, leading to corrections to the bulge's projection on the binary axis that are quadratic in the diffusion coefficients. Consequently, as we expect, this introduces quadratic corrections to the distortion's primary effect on the gravitational wave signal.}
and suppresses the perturbation amplitude by a factor
$\sqrt{1+\frac{1}{m^2\Omega^2 \tau^2_{\rm diff,\,\bf k}}}$. This factor
is of the second order in diffusion coefficients when $1/(m\Omega)\ll \tau_{\rm diff,\,\bf k}$ and is large  when $1/(m\Omega)\gg \tau_{\rm diff,\,\bf k}$.
The former is the case for the f-modes, while the latter precisely describes the case for the 
$m=\pm 2$ g-modes at large orbital separations.


\section{Numerical values of various quantities}\label{AppE}

Here the interested reader may find numerical values of various quantities entering the discussion of the main text.

\begin{figure}
\center{
\includegraphics[height=0.33\textheight]{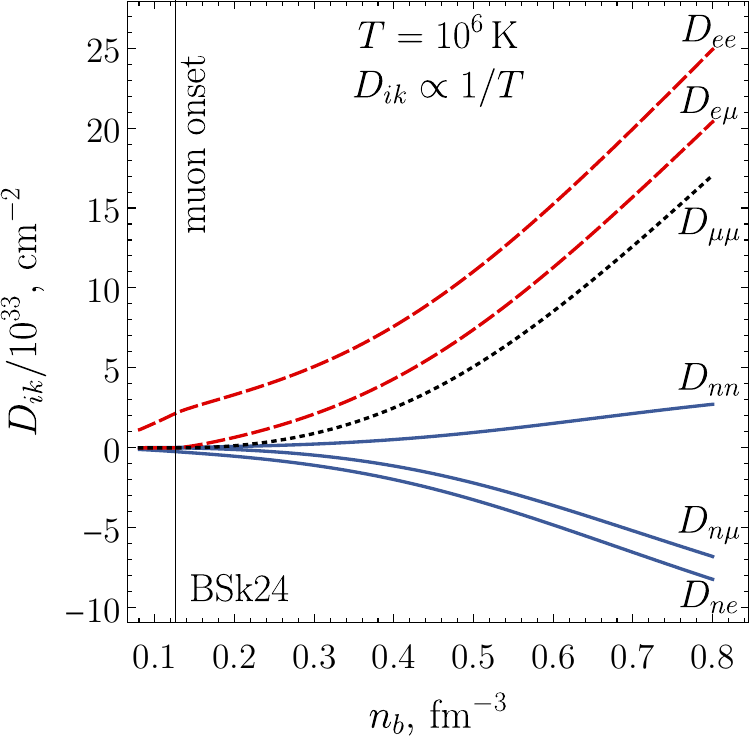} \hspace{0.5cm}}
\caption{
Diffusion coefficients $D_{ik}=D_{ki}$ as functions of the baryon number density $n_b$ at {\it local} temperature $T=10^6 \, \rm K$. They can be recalculated to other $T$ using the approximate scaling $D_{ik}\propto 1/T$.}
\label{Fig:DikNB}
\end{figure}

\begin{figure}
\center{
\includegraphics[height=0.33\textheight]{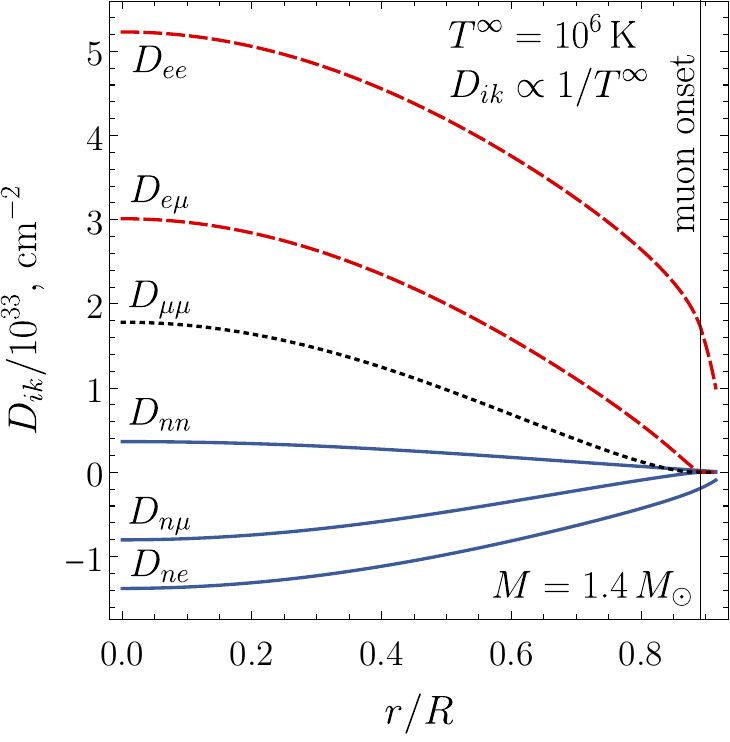}}
\caption{
Diffusion coefficients calculated at {\it redshifted} temperature $T^\infty=10^6 \, \rm K$ as functions of the normalized radial coordinate $r/R$ for $M=1.4M_\odot$ stellar model. They can be recalculated to other $T^\infty$ using the approximate scaling $D_{ik}\propto 1/T^\infty$.
}
\label{Fig:Dik}
\end{figure}

First, we calculate the diffusion coefficients $D_{ik}$ using the algorithm described in Appendix C of Ref.\, \cite{dgs20}. 
To do that, as mentioned in the main text, 
we treat 
the 4-component $npe\mu$-matter with strongly superconducting protons as 
effectively
the 3-component $ne\mu$-matter. 
In this case diffusion is described by the coefficients $D_{nn}$, $D_{ne}=D_{en}$, $D_{n\mu}=D_{\mu n}$, $D_{ee}$, $D_{e\mu}=D_{\mu e}$ and $D_{\mu\mu}$. 
In Fig.\ \ref{Fig:DikNB} we present these coefficients as functions of the baryon number density $n_b$, while Fig.\ \ref{Fig:Dik} shows their values as functions of the normalized radial coordinate $r/R$. Our numerical calculations reveal that $D_{ik}$ scale with the temperature approximately as $D_{ik}\propto 1/T^\infty$.


%
%

\begin{figure}
\center{\includegraphics[width=0.5\linewidth]{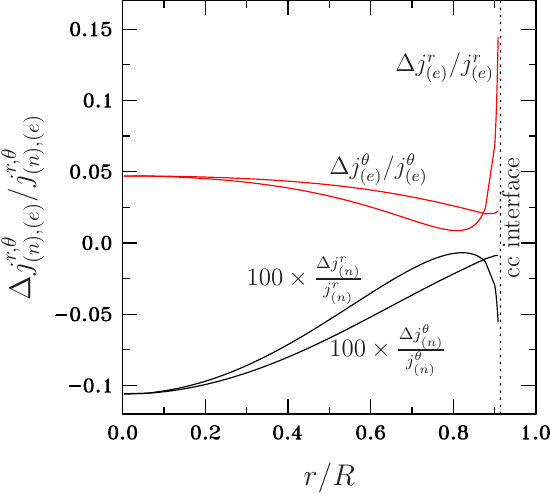}}
\caption{Solid lines show ratios $\Delta j^{r,\theta}_{(n),(e)}/j^{r,\theta}_{(n),(e)}$ as functions of normalized radial coordinate calculated at $\nu_{\rm GW}=1.5\,\rm Hz$ and corresponding stellar temperature (see Fig.\ \ref{Tpic}) for $l=2$ $m=\pm 2$ f-mode of $M=1.4M_\odot$ stellar model. Vertical dots show core-crust interface.}
\label{Fig:Deltaj}
\end{figure}

Next, Fig.\ \ref{Fig:Deltaj} shows the ratios $\Delta j^{r,\theta}_{(n),(e)}/j^{r,\theta}_{(n),(e)}$ %
\footnote{More accurately,
the ratios of the amplitudes of $\Delta j^{r,\theta}_{(n),(e)}$ and $j^{r,\theta}_{(n),(e)}$, since they have a phase shift in their time dependence.} 
calculated at $\nu_{\rm GW}=1.5\,\rm Hz$  for the $l=2$ $m=\pm 2$ f-modes. The ratios $\Delta j^{r,\theta}_{(\mu)}/j^{r,\theta}_{(\mu)}$ are not shown, since they are indistinguishable from $\Delta j^{r,\theta}_{(e)}/j^{r,\theta}_{(e)}$.%
\footnote{This is a result of an approximate relation between the diffusion coefficients: $D_{\mu i}\approx n_\mu/n_e D_{e i}$.} 
For the approach adopted in the paper to be applicable, the plotted ratios have to be small.
Figure \ref{Fig:Deltaj} implies that while $\Delta j^{r,\theta}_{(n)}/j^{r,\theta}_{(n)}$ are indeed small, $\Delta j^{r,\theta}_{(e,\mu)}/j^{r,\theta}_{(e,\mu)}$ can be regarded as small quantities with less accuracy%
\footnote{The strong difference between $\Delta j^{r,\theta}_{(n)}/j^{r,\theta}_{(n)}$ and $\Delta j^{r,\theta}_{(e,\mu)}/j^{r,\theta}_{(e,\mu)}$ has a clear 
explanation.
Consider, for simplicity,  $npe$-matter. 
Using Eq.\ (\ref{Dmk}) and symmetry of the matrix $D_{ik}$, one can obtain $\frac{\Delta j^{r,\theta}_{(e)}}{j^{r,\theta}_{(e)}}=-\frac{n_n \mu_n}{n_e \mu_e}\frac{\Delta j^{r,\theta}_{(n)}}{j^{r,\theta}_{(n)}}$. The ratio $\frac{n_n \mu_n}{n_e \mu_e}$ is a large quantity with typical values $\sim 100$ throughout the core.}.
However, one should keep in mind that these ratios decrease during the binary evolution (since they are $\propto \nu_{\rm GW}^{-1}(T^\infty)^{-2}$ and both $\nu_{\rm GW}$ and $T^\infty$ increase with time), while Fig.\ \ref{Fig:Deltaj} is plotted for the earliest moment assumed to be observed, corresponding to $\nu_{\rm GW}=1.5\,\rm Hz$.


\begin{figure}
\center{\includegraphics[width=0.5\linewidth]{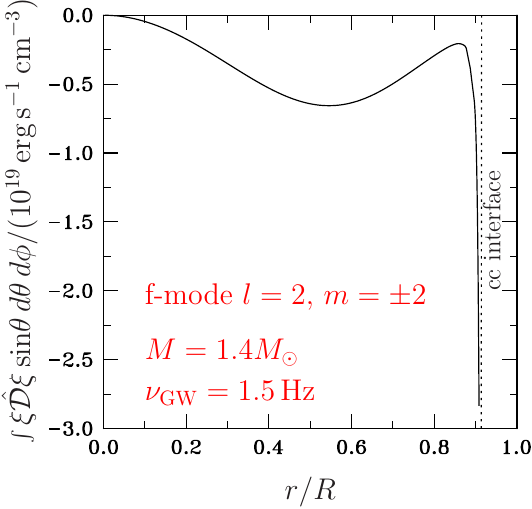}}
\caption{Angle-integrated perturbation energy dissipation rate $\int \bs \xi \hat{\mathcal{D}} \bs \xi\,{\rm sin}\theta\,d\theta\,d\phi$ as a function of the normalized radial coordinate calculated at $\nu_{\rm GW}=1.5\,\rm Hz$ and corresponding stellar temperature (see Fig.\ \ref{Tpic}) for the $l=2$ $m=\pm 2$ f-modes of $M=1.4M_\odot$ stellar model. Vertical dots show the crust-core interface.}
\label{Fig:Edot}
\end{figure}

In Fig.\ \ref{Fig:Edot} we show the angle-integrated perturbation energy dissipation rate $\int \bs \xi \hat{\mathcal{D}} \bs \xi\,{\rm sin}\theta\,d\theta\,d\phi$ as a function of the normalized radial coordinate $r/R$. The calculation is performed for the $M=1.4M_\odot$ stellar model at $\nu_{\rm GW}=1.5\,\rm Hz$ and the corresponding stellar temperature (see Fig.\ \ref{Tpic}). Only the $l=2$ $m=\pm 2$ f-modes are accounted for. As one can see, the dissipation occurs most efficiently near the crust-core interface.


\begin{figure}
\center{\includegraphics[width=0.5\linewidth]{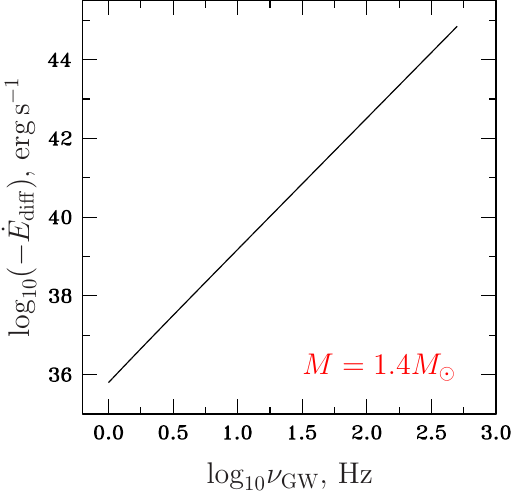}}
\caption{Heating rate $-\dot E_{\rm diff}$ as a function of GW frequency calculated for $M=1.4M_\odot$ stellar model and stellar temperature from  Fig.\ \ref{Tpic}.}
\label{Fig:EdotTotal}
\end{figure}

Fig.\ \ref{Fig:EdotTotal} shows the heating rate of the $M=1.4M_\odot$ NS as a function of the GW frequency. We see that in logarithmic scale this dependence is linear.
This linearity follows from the evolution laws for $\nu_{\rm GW}$ and $T^\infty$, $\dot \nu_{\rm GW}\propto \nu_{\rm GW}^{11/3}$ and $\dot T^\infty \propto (T^\infty)^{-3}\nu_{\rm GW}^{4}$ (see Eqs.\ \ref{Omegaprop} and \ref{Tprop}), which imply $T^\infty\propto \nu_{\rm GW}^{1/3}$. As a result, we arrive at $\dot E_{\rm diff}\propto D^{-6} (T^\infty)^{-2}\propto\nu_{\rm GW}^4 (T^\infty)^{-2}\propto \nu_{\rm GW}^{10/3}$.


\begin{figure}
\center{\includegraphics[width=0.5\linewidth]{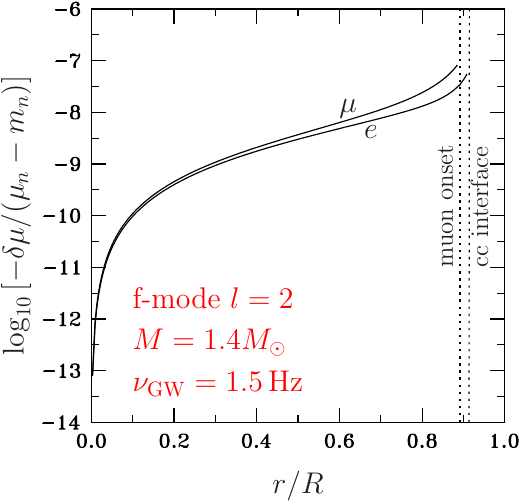}}
\caption{Normalized amplitudes of chemical potential imbalances as functions of the normalized radial coordinate calculated at $\nu_{\rm GW}=1.5\,\rm Hz$ for $l=2$ f-mode  and $M=1.4M_\odot$ stellar model. Angular dependence of $\delta \mu$ ($\propto Y_{lm}$) is factored out here.}
\label{Fig:deltamu}
\end{figure}

Finally, Fig.\ \ref{Fig:deltamu} shows the radial profiles of the logarithms of two (electron and muon) chemical potential imbalances $\delta \mu$ ($\delta \mu\equiv \mu_n-\mu_p-\mu_l$, where $l=e$ or $\mu$ for the electron and muon chemical potential imbalance, respectively). We normalize these chemical imbalances to the ``nonrelativistic'' neutron chemical potential $\mu_n-m_n$ (chemical potential of neutrons measured from the bare mass of the neutron).
The normalized chemical imbalances characterize the degree of the deviation
from chemical equilibrium, and, since diffusive dissipation can be expressed in terms of chemical potential imbalances, they also show the degree of 
deviation from the diffusion equilibrium. 


%


\end{document}